# Correlative Synchrotron X-ray Imaging and Diffraction of Directed Energy Deposition Additive Manufacturing

Yunhui Chen[1,2], Samuel J. Clark[1,2], David M. Collins[3], Sebastian Marussi[1,2], Simon A. Hunt[4], Danielle M. Fenech[5], Thomas Connolley[6], Robert C. Atwood[6], Oxana V. Magdysyuk[6], Gavin J. Baxter[7], Martyn A. Jones[7], Chu Lun Alex Leung[1,2], Peter D. Lee[1,2]

[1] Mechanical Engineering, University College London, Torrington Place, London WC1E 7JE, UK

[2] Research Complex at Harwell, Rutherford Appleton Laboratory, Oxfordshire OX11 0FA, UK

[3] School of Metallurgy and Materials, University of Birmingham, Edgbaston, Birmingham B15 2TT, UK

[4] Department of Materials, University of Manchester, Oxford Rd, Manchester M13 9PL, UK

[5] Department of Physics, Cavendish Laboratory, University of Cambridge, JJ Thompson Avenue, Cambridge, Cambridgeshire CB3 0HE, UK

[6] Diamond Light Source, Harwell Campus, Oxfordshire, OX11 0DE, UK

[7] Rolls-Royce plc, PO Box 31, Derby, DE24 8BJ, UK

## Abstract

The governing mechanistic behaviour of Directed Energy Deposition Additive Manufacturing (DED-AM) is revealed by a combined *in situ* and *operando* synchrotron X-ray imaging and diffraction study of a nickel-base superalloy, IN718. Using a unique process replicator, real-space phase-contrast imaging enables quantification of the melt-pool boundary and flow dynamics during solidification. This imaging knowledge informed precise diffraction measurements of temporally resolved microstructural phases during transformation and stress development with a spatial resolution of 100 µm. The diffraction quantified thermal gradient enabled a dendritic solidification microstructure to be predicted and coupled to the stress orientation and magnitude. The fast cooling rate entirely suppressed the formation of secondary phases or recrystallisation in the solid-state. Upon solidification, the stresses rapidly increase to the yield strength during cooling. This insight, combined with IN718's large solidification range suggests that the accumulated plasticity exhausts the alloy's ductility, causing liquation cracking. This study has revealed additional fundamental mechanisms governing the formation of highly non-equilibrium microstructures during DED-AM.

# 1. Introduction

Laser Additive Manufacturing (LAM)[1] is a highly versatile and flexible manufacturing technology, enabling layer-by-layer fabrication of complex geometries. It is now transforming modern manufacturing[2], especially in metallurgical sectors[3]. Directed Energy Deposition Additive Manufacturing (DED-AM)[4], which deposits powder or wire feedstock through a nozzle and melts it with a laser, is one of the most cost-effective and versatile LAM methods due to its capability for producing large near-net-shape freeform components and its utility for the repair of high-value components in aerospace[5], biomedical[6] and automotive industries[7]. However, rapid solidification during the DED-AM process results in several technical challenges including the creation of significant residual stresses[8] and the formation of undesirable microstructural features such as pores, cracks or large epitaxial grains[9]; this presently restricts the widespread industrial application of DED-AM for producing safety-critical components. To overcome these limitations, an increased understanding of the underlying transient physics during the manufacturing process is required. Trial-and-error methods have been used to establish the relationship between processing conditions, resultant microstructures and properties; however, this approach is unsatisfactory if we are to understand and optimise the competing transient mechanisms that govern the build quality and subsequent material and component performance.

*In situ* and *operando* high-speed X-ray investigations of LAM have proved enormously successful for revealing previously unseen transient laser-induced phenomena including melt pool dynamics[10,11], microstructural feature formation[12] and phase evolution[13–15]. Synchrotron X-ray imaging has been shown to effectively capture the laser-matter interactions[16] and the underlying physics in laser powder bed fusion (LPBF)[17–20]. However, much less attention has been given to synchrotron imaging of DED-AM. Due to the larger length scales of industrial DED-AM deposits leading to poor X-ray transmission, X-ray studies are more challenging. Despite this, it remains highly desirable to quantify and understand optically opaque metallic samples with high spatial and temporal resolution. Such insights have become possible with high-flux, high energy third-generation synchrotron radiation sources[21] that enable fast (millisecond to microsecond) X-ray imaging and diffraction of the laser-matter interactions. Wolff et al [21] simulated some aspects of the blown powder process by using a piezo-driven vibration-assisted powder delivery system to induce a gravitational flow of powders from a syringe-needle. Chen et al[22-23] performed the *in situ* X-ray imaging on industrial scale DED-AM and revealed the materials behaviour differences between titanium alloys and stainless steel. However, diffraction analysis of DED-AM remains unexplored.

Recently, fast synchrotron X-ray diffraction methods were used to reveal microstructural evolution and thermal gradients during the LPBF process, providing information missing from X-ray imaging[15]. Zhao *et al.*[14] demonstrated synchrotron diffraction could be used to capture phase transformations in LPBF*.* Thampy *et al.*[22] further estimated the subsurface cooling rate during LPBF using diffraction peak position shifts. Hocine *et al.*[13,23] implemented an ultra-fast diffraction detector to estimate phase transformations, cooling rates and residual stresses using peak intensity changes in different printing parameters in LPBF. Although the temperature and phase transformation were monitored from the diffraction patterns with a high temporal resolution but low spatial resolution, quantitative analysis of phase fraction, stress and liquid fraction were not determined. Further, none of these studies examined DED-AM.

The material investigated in this study was nickel-base superalloy IN718. It has excellent high-temperature performance and corrosion resistance, and is widely applied for safety-critical components such as turbine discs [24] in aerospace, marine and power generation gas turbines. IN718 is also a commonly used superalloy for LAM. However, LAM introduces high thermal stresses which originate from the high elastic modulus and thermal expansion coefficient of IN718[25]. The formation of eutectic compounds and elemental segregation at the grain boundaries and/or interdendritic regions from the addition of Ti or Nb increases hot cracking susceptibility[26]. As a result, it can be difficult to deposit crack free components using LAM.

The high-temperature strength of IN718 is attributable to a fine dispersion of stable $L1_2$ structure $\gamma'$ precipitates and metastable $D0_{22}$ structure $\gamma''$ precipitates. As the formation of these desirable phases are solid-state diffusion controlled, the high thermal gradient and rapid cooling rates may suppress their formation during the LAM process. Further secondary phases including carbides, Laves and δ phases are often found in IN718 during LAM[27–30], which are known to deplete the elements needed to form the strengthening phases during post-processing. Significant research has been conducted to control the microstructure and remove undesirable features by post-heat treatments[31] and hot isostatic pressing. Understanding the as deposited microstructure and detrimental feature formation during AM is critical for the development of *in situ* and post build heat-treatment regimes.

In this work, we perform combined temporally and spatially resolved X-ray imaging and diffraction of the DED-AM process using a unique AM process replicator with capabilities that directly scales to replicate industrial process parameters. The combination of X-ray imaging and diffraction provides a holistic in-depth understanding of DED-AM including the melt pool dynamics, solidification sequence and undesirable microstructural feature formation captured *in situ* and operando. This investigation first enabled quantification of key features via Xray imaging, guiding diffraction quantification of temperature, strain and phases, all spatially mapped across the weld pool and the surrounding heat-affected region.

## 2. Material and Methods

### 2.1 Materials

The material used was a PREP IN718 powder (Timet, USA) with a size distribution of 45 - 90 µm and with median particle diameter, $D_{50}$, of 70 µm. The chemical compositions of IN718 powder used in this study are shown in Table 1.

Table 1. Chemical composition of IN718 powder used (wt %)

| **Ni** | **Cr** | **Fe** | **Nb** | **Mo** | **Ti** | **Al** | **Co** |
|---|---|---|---|---|---|---|---|
| 50.0-55.0 | 17.0-21.0 | Bal | 4.75-5.5 | 2.8-3.3 | 0.65-1.15 | 0.2-0.8 | <1.0 |
| **C** | **Mn** | **Si** | **P** | **S** | **B** | **Cu** | **O, N** |
| <0.08 | <0.35 | <0.35 | <0.015 | <0.015 | <0.006 | <0.3 | <250 ppm |

### 2.2 *In situ* and *operando* synchrotron X-ray imaging and diffraction of DED-AM

We performed *in situ* and *operando* X-ray imaging and diffraction on the I12: Joint Engineering, Environmental, and Processing (JEEP) beamline[21] at the Diamond Light Source to capture the transient phenomena and the underlying physics during the DED-AM of IN718. The Blown Powder Additive Manufacturing Process Replicator (BAMPR) has been developed to faithfully replicate a commercial DED-AM system with an industrial laser power density (up to 6366 W mm$^{-2}$). that can be integrated into synchrotron beamlines (see **Supplementary Video 1**)[32]. The traverse speed of the sample stage in both cases was controllable in the range 1 - 5 mm s$^{-1}$ to enable continuous track formation. **Further details can be found in Methods and Supplementary Information and in ref. [23].**The resulting time-series radiographs and Debye-Scherrer diffraction patterns were subsequently analysed to quantify the mechanisms and dynamics of the DED-AM process.

### 2.3 Marangoni flow tracing using W tracer

Tungsten (W) particles ($D_{50}$ of 50 µm) were selected to match spatical resolution of the imaging detector and employed as tracers to visualise the flow in the melt pool by observing their spatiotemporal distribution on the separate set of imaging experiement. 4 wt% of W particles were pre-mixed in the IN718 powder feedstock. During the DED-AM process, the particles were seeded into the melt pool and used as fiducial markers to track the melt flow. As the diameter of W particles are similar to IN718 powder feedstock, their influence to the flow dynamics are negligible. Due to the high melting point of W, the particles remained solid before being engulfed by the solidification front and encapsulated in the melt track. W has a significantly higher X-ray attenuation compared to IN718, thus they appeared darker than the surrounding liquid metal in the acquired radiographs. A maximum temporal resolution (0.2 ms) was exploited to capture the fast dynamics in the melt pool. The ImageJ Plugin TrackMate[33] was used to trace the trajectories of the W particles to analyse the dynamic flow directions and velocities.

### 2.4 Image processing and quantification

All acquired radiographs were processed using ImageJ[34] and Matlab©. The acquired images were corrected to form a flat-field corrected (FFC) image by dividing by the mean average of 100 flat field images[11]. To increase the image contrast and signal-to-noise ratio, a local-temporal background subtraction was applied to reveal low-intensity features in the X-ray images following equation:

$$\mathrm{LTBS} = \frac{FFC}{I_{\mathrm{lavg}}} \quad (1)$$

where LTBS is the local-temporal background-subtracted image, *FFC* is the flat field corrected image, and $I_{lavg}$ is a local average of 50 of the nearest neighbour images (25 before and 25 after).

### 2.5 Diffraction processing and quantification

Images of the Debye-Scherrer diffraction patterns were radially integrated using DAWN[35] into 36 azimuthal sectors. All subsequent peak fitting and data analysis were performed using Matlab© and python. To monitor the phase transformations during DED-AM, a pseudo-Voigt function was fitted to individual reflections, recording d-spacing and integrated peak intensity as well as their associated fitting error. Fourier smoothing technique was implemented to fit the caked patterns, which are used to determine the strains. The fitted peaks and their fitting errors were used to assist phase identification, secondary phase volume fraction, temperature and stress analysis. **Full details of the methodology are described in the Supplementary Methods (2.1-2.4).**

# 3. Results

### 3.1 *In situ* and *Operando* X-ray imaging and diffraction

Synchrotron X-ray imaging and diffraction on the nickel-based superalloy was performed at the I12:JEEP beamline[21] to capture the transient phenomena *in situ* and *operando* during DED-AM. The imaging experiments were used to determine the detector positions for the diffraction measurements.

During the *in situ* imaging experiments, the X-ray beam was attenuated by the substrate, powder and the deposited melt track and a PCO-Edge CCD camera enables the acquisition of the resulting radiographic video, as shown in Figure 1b. These radiographs record the time-resolved evolution of the multi-layer melt track morphology of a DED-AM build using IN718 (Figure 1b and Supplementary Video 1). The framerate and exposure time used in the high-resolution imaging experiment were 200 fps and 0.0049 s, respectively. The frame rate was selected to optimise the signal to noise ratio whilst being a sufficiently fast temporal resolution to capture key phenomena in DED-AM's large melt pool (10x bigger than LPBF in each direction)[10], requiring a large field of view (FoV) and high penetration through the thick nickel superalloy track. These values were changed to a 5000 fps framerate and 0.00019 s exposure time in the separate experiments used to capture the fast Marangoni convection in the melt pool. The synchrotron imaging conditions were carefully tailored to optimise the trade-off between signal to noise, spatial and temporal resolution. **Further details of the acquisition conditions can be found in the Supplementary Information(1.3).**

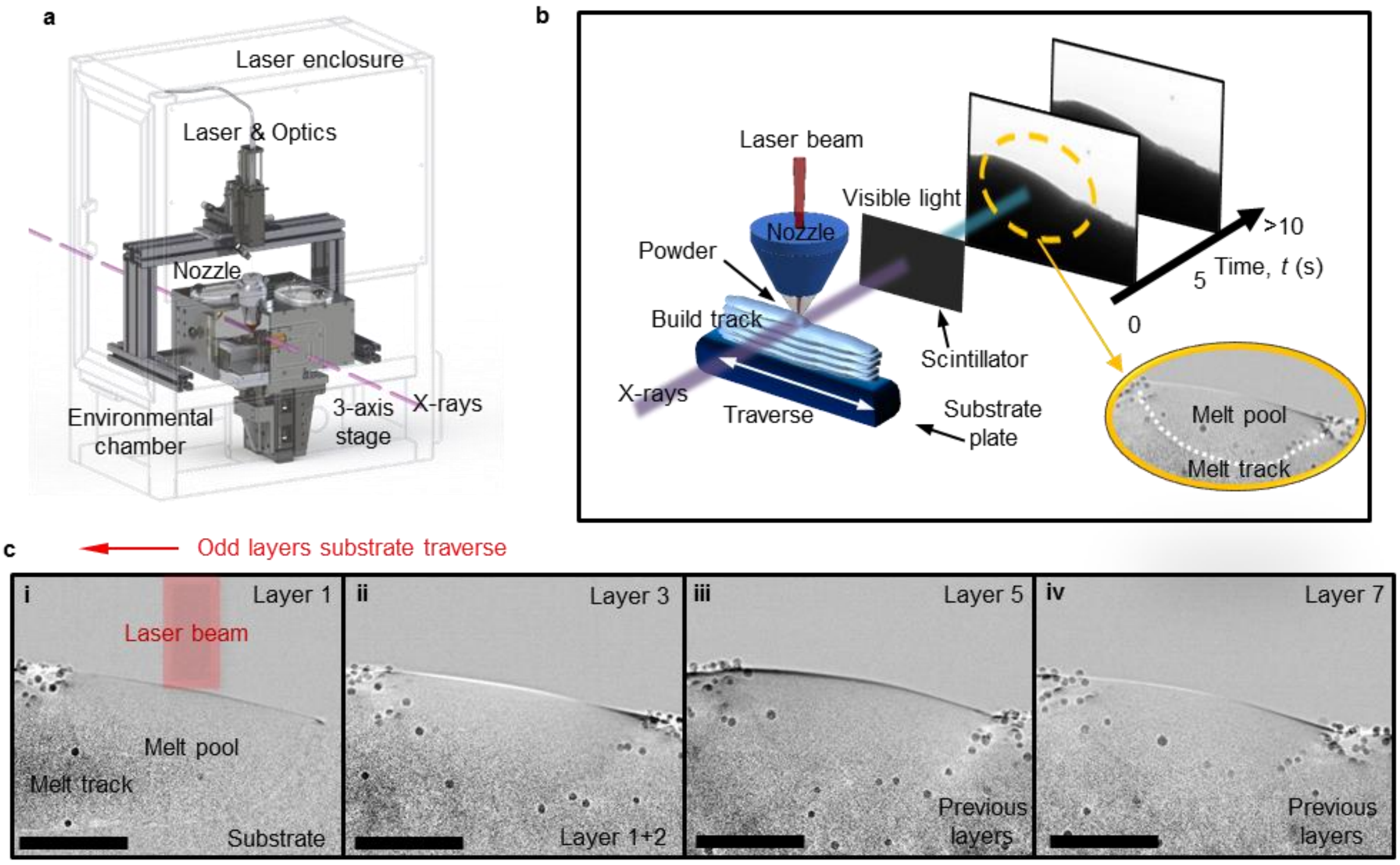

*Figure 1. In situ X-ray imaging quantification of DED-AM IN718: (a) the Blown Powder Additive Manufacturing Process Replicator (BAMPR) designed to reproduce the operation of a commercial DED-AM system. The system is encapsulated within a Class I laser enclosure and comprises an inert environmental chamber, a high precision 3-axis platform, an industrial coaxial DED nozzle, and a laser system (see Suppl. video 1). (b) Schematic of the in situ X-ray imaging of DED-AM process. The synchrotron X-rays are attenuated by the deposited materials and then are converted to visible light with a scintillator to capture as a radiographic video on a fast CCD camera (not shown). The resulting videos (Supplementary Video 2) reveal the underlying mechanism and dynamics of the process. (c) Time-series radiographs acquired during DED-AM of a multilayer thin wall melt track of IN718 under P = 200 W, v = 1 mm s$^{-1}$ captured at 200 fps. Scale bar = 500 µm. Four filtered radiographs were selected and as representative of the multilayer thin wall building process of (i) layer 1, (ii) layer 3, (iii) layer 5 and (iv) layer 7, respectively.*

The single hatch thin-wall melt tracks were deposited in an alternating bi-directional strategy up to 10 layers in height. Figure 1c shows a filtered time-series of radiographs taken during a multilayer deposition process. The substrate was traversing from the right to the left in the figure shown for odd layer builds (see **Supplementary Video 2)**. The laser beam is shown to melt the substrate plate surface and consolidate the powder particles into a melt pool which forms a continuous deposited track. The melt pool area was identified by the solid-liquid boundaries. The molten pool surface is observed to oscillate during the building process, which can be attributed, in part, to the recoil of the molten surface as powder particles impact with it. An additional cause for this oscillation is the presence of a thermal gradient across the melt pool which induces the Marangoni flow. The thermal gradient across the molten pool is further complicated by the quenching effect of cooler powder particles being incorporated into the melt pool. Thus, the dynamics of flow in the molten pool result from the complex interplay between the Marangoni flow and particle incorporation. A smooth surface finish and some un-melted powder particles can be seen on the newly formed melt track in IN718 in all layers of the build. In the subsequent layers of the build, the laser beam partially re-melts the previous build layers and a new layer is formed on top. The surface finish and the melt pool geometries are seen to be consistent in all layers of the build and no apparent undesirable microstructural features such as porosity or cracking were observed in the radiographs.

*In situ* diffraction data were acquired using a Pilatus 2M CdTe 2D area detector, enabling full Debye-Scherrer diffraction patterns to be collected in transmission. An X-ray energy of 70 keV was used and calibrated with a $CeO_2$ standard. The experimental set-up is shown in Figure 2a. Our study enabled the acquisition of the diffraction evolution during a thin wall build (up to 5 layers) by systematically moving the area detector in directions orthogonal to the laser beam, mapping the melt pool and melt track area. The wall roughness is small, and the X-ray transmission through the thickness of the melt track (~1.5 mm) is uniform, which is evidence by Figure 2(b) & (c). Two mapping regions were selected; the first being a high spatial accuracy map of the melt pool region to capture the solidification phenomena. This strategy consisted of a map of 6×7 (row x columns) data points with a 100 μm step size both horizontally and vertically, perpendicular to the build and laser directions. Secondly, a much larger region of the melt track area extending well beyond the rear of the pool was mapped to understand the stress evolution in the newly formed melt track. This strategy contains a map of 3 × 20 (row x columns) data points with a 100 μm step size horizontally and a 200 μm step size vertically. The beam size was 100 × 100 $\mu m^2$ in all experiments.

As the laser beam and the resulting melt pool were stationary in respect to the detector during the substrate traverse, one diffraction pattern was recorded at each mapping point with an exposure time of 6.67 s, effectively capturing the averaged signal from a volume of 100 μm × 6.67 mm × wall thickness. Figure 2d shows the diffraction intensity as a function of time for the 1st layer of the DED-AM build. The data collection procedure for the other layers was similar. The data collected are averaging information through wall thickness. The diffraction geometry enabled the identification of the {111}, {200}, {220}, {311} and {222} reflections of the $\gamma$ phase, {111}, {200}, {220}, {311} reflections of the MC carbide phase TiC. The (110), (310), (211), (120), (400), (410), (510), (030), (321), (230), (520) and (022) reflections of a Laves phase, can be found at the end of the DED-AM process.

Prior to the material deposition, at *D* = 0 mm, only the $\gamma$ phase reflections can be seen from the substrate. The initiation of laser melting is indicated by the formation of liquid. Once the laser moved past, at *D* = ~1.2 mm, the molten material started to solidify, indicated by the emergence of sharp reflections associated with a crystal phase. The line profiles of the $\gamma$ phase gradually appeared and the intensity increased during the solidification process. Reflections of the secondary phases including Laves and Carbides appeared soon after, revealing the solidification sequence during the cooling process of DED-AM. The full sequence was identified to be: Liquid → Liquid + $\gamma$ → Liquid + $\gamma$ + MC → Liquid + $\gamma$ + MC + Laves → $\gamma$ + MC + Laves. The thermal field in the region of the weld pool was also inferred from the diffraction measured d-spacing changes observed in Figure 2d. In the volume probed, the temperature firstly increased before the melting process, indicated by a unit cell expansion, and decrease during solidification, indicated by a unit cell contraction.

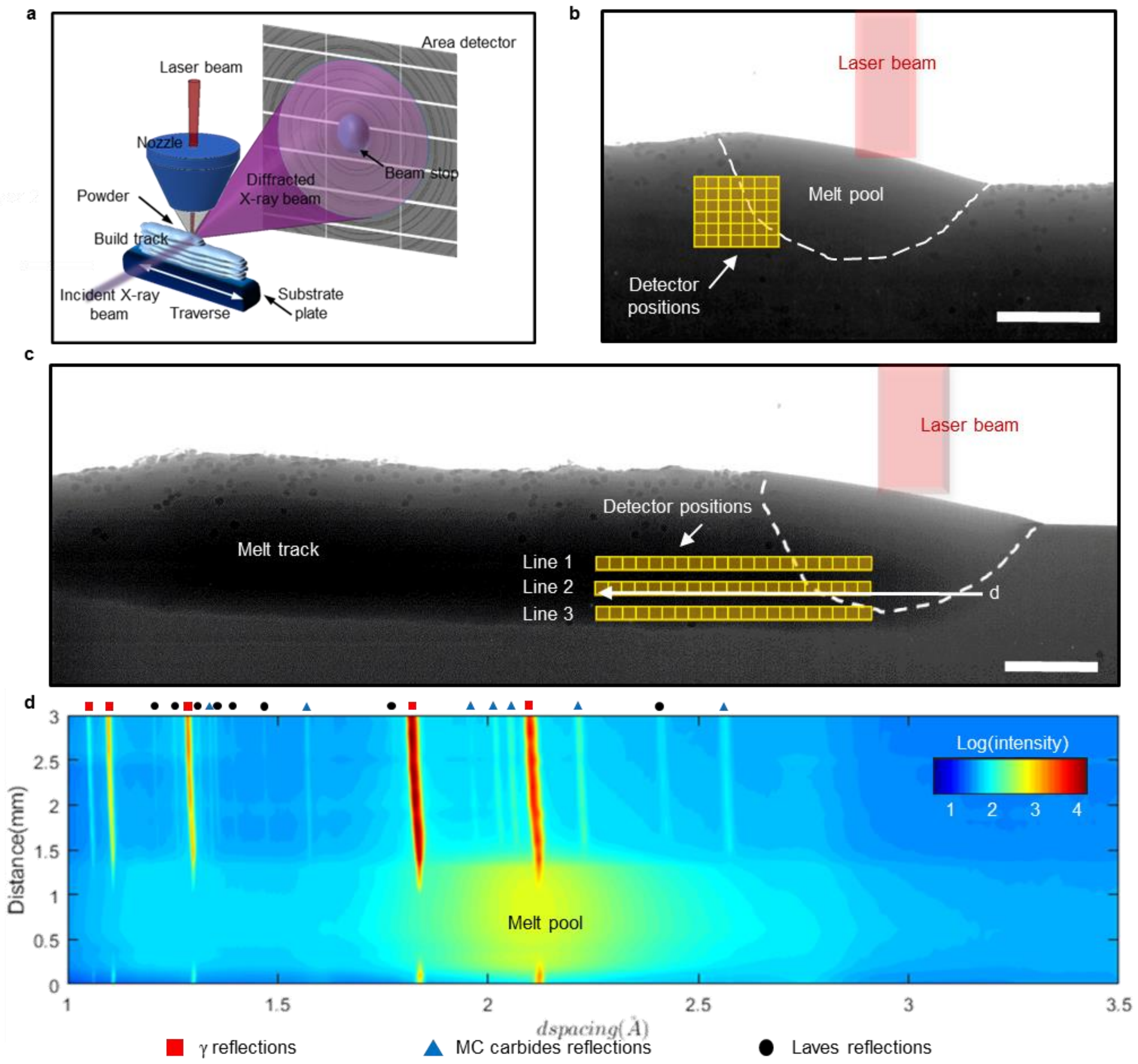

*Figure 2. In situ X-ray diffraction of DED-AM IN718: (a) Schematic of the in situ and operando X-ray diffraction of DED AM process. The synchrotron X-rays are attenuated by the deposited materials and the diffracted X-rays are recorded by a large 2D area diffraction detector. (b) Melt pool mapping strategy used in this study, consisting of 6×7 mapping matrix. (c) Melt track mapping strategy used in this study, consisting of 3×20 mapping matrix for the melt track area. (d) Phase evolution during DED-AM of a single track, shown as intensity as a function of distance and d-spacing. The scanning strategy is shown in (c) following the arrow 'd'. The indexing of the identified peaks can be found in Supplementary Results(3.2).*

### 3.2 Flow dynamics during DED-AM

The flow dynamics in the molten pool are a complex interplay between the fluid flow coupled with the damping effect of powder particle incorporation. The laser-induces a high thermal gradient across the melt pool which induces Marangoni flow; this is further complicated by the quenching effect of cooler powder particles melting as they are incorporated into the melt pool.

Tungsten tracers were added to quantify the melt pool flow patterns in the multi-layer DED-AM build and the flow pattern observed, showing that the melt pool shape is largely determined by the flow characteristics (Figure 3 and **Supplementary Video 3**). The trajectories of flowing W tracer particles indicate radial outward flow paths from the centre of the melt pool in the upper part and inward flow path at the bottom, as shown in the schematic (Figure 3a). The melt pool is divided into three regions: 1. Region A, where a clockwise flow is seen at the right side of the melt pool (Figure 3b); 2. Region B, an anti-clockwise flow is observed at the left side of the melt pool (Figure 3c); 3. Region C, parallel to the X-ray beam direction flow (in-and-out of page direction in Figure 3a) in the centre of the melt pool, which exhibits up-and-down behaviour in 2D projection. The coloured lines indicate the trajectory and direction of the tracer particles. Radiographs recorded during the experiments are the 2D projection of the 3D flow, and the schematic of the actual 3D flow is shown in Figure 3d. The projected average speed of the tracers flowing in the melt pool was measured to be ~ 2 mm $s^{-1}$ in all layers of the build (see **Supplementary Results(3.1)** for instantaneous tracer velocities).

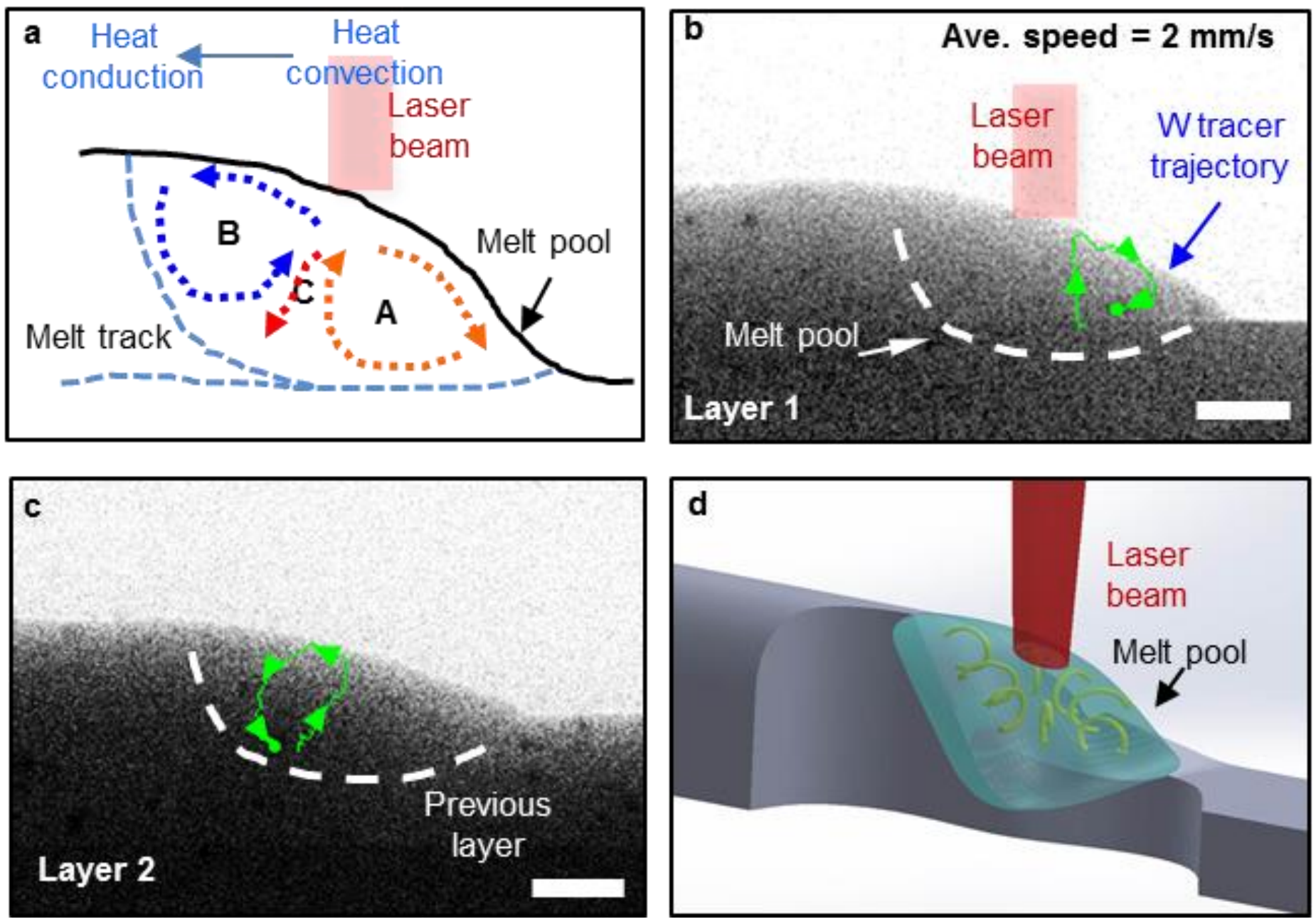

*Figure 3. In situ and operando X-ray imaging quantification of Marangoni flow during DED-AM IN718 using W tracers. The synchrotron X-rays attenuation differences between IN718 and W create imaging contrast used to measure the speed and direction of the Marangoni convection in the DED-AM melt pool. A video showing the W tracers following the Marangoni flow can be found in Supplementary Video 3. (a) Schematic showing the Marangoni convection in DED-AM. The melt pool is divided into three regions in 2D projections recorded by synchrotron X-ray radiography. (b) A time-series radiograph revealing the trajectory of a W tracer in region A in the melt pool following Marangoni convection during first-layer of build under P = 200 W, v = 1 mm s$^{-1}$ captured at 5000 fps. Scale bar = 500 µm. (c) Time series radiographs revealing the trajectory of a W tracer in region B in the melt pool following Marangoni convection during second-layer of the build. (d) Schematic of the 3D Marangoni convection in DED-AM.*

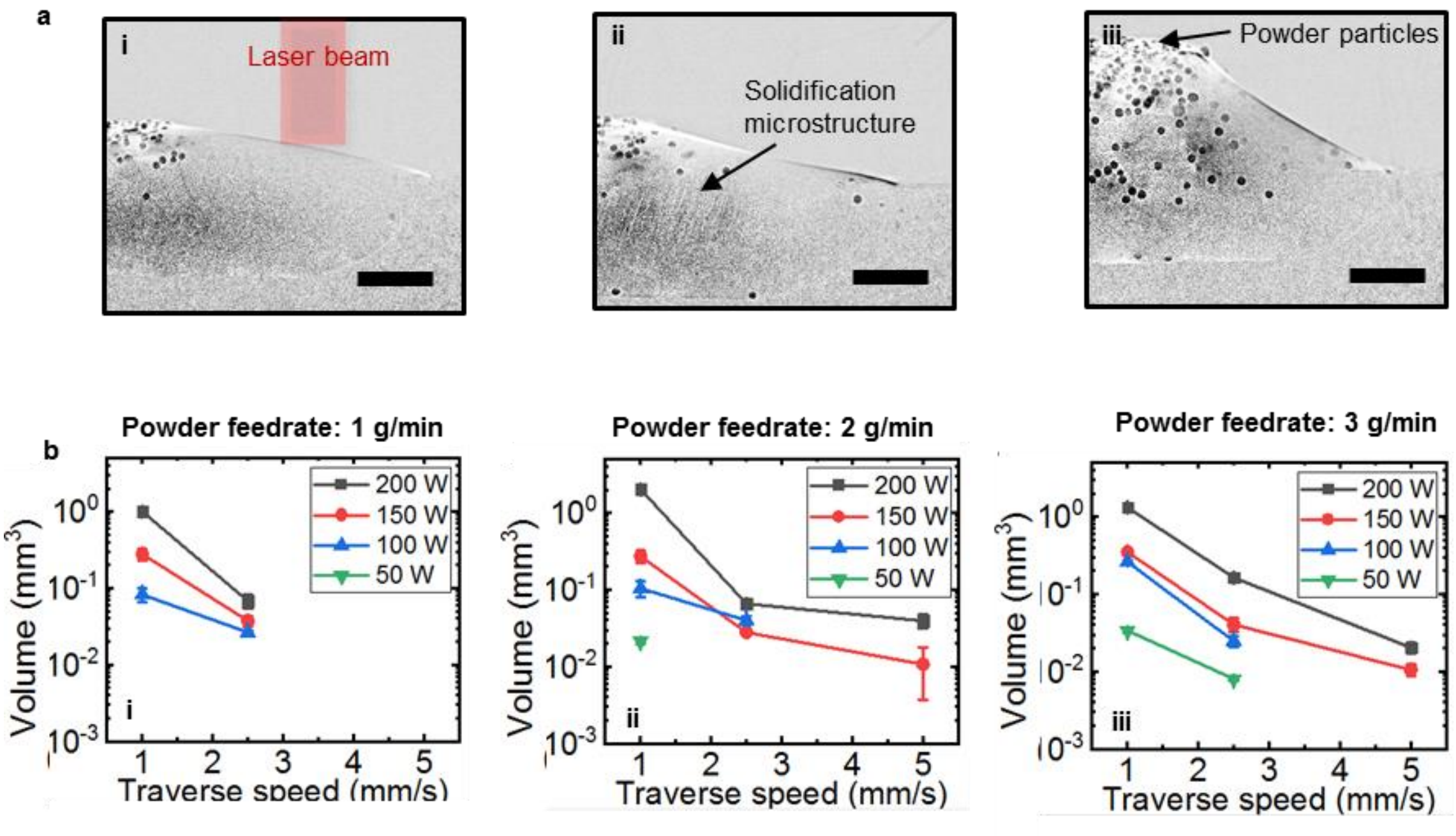

*Figure 4. In situ synchrotron X-ray imaging melt pool quantification of DED-AM IN718. Melt pool volumes and morphologies for varying powder feedrate, laser power and traverse speed with (a) representative radiographs acquired during DED-AM of a multilayer thin wall melt track of IN718 under P = 200 W, v = 1 mm s$^{-1}$ , (i) f = 1g min$^{-1}$, (ii) f = 2g min$^{-1}$, (iii) f = 3g min$^{-1}$, captured at 200 fps. Scale bar = 500 µm. (b) Effective melt pool volumes measurements across powder feedrate, laser power and traverse speed. Error bars indicate SD.*

The melt pool volumes and geometries across operating parameters were quantified from the synchrotron X-ray radiography (Figure 4). The melt pool volume is plotted as a function of the powder feedrate, laser power and traverse speed. Radiographs of the typical pool profiles with powder feedrate of 1 g min$^{-1}$, 2 g min$^{-1}$and 3 g min$^{-1}$ are shown in Figure 4a i-iii, respectively. The melt pool volume increases with increasing powder feedrate and laser power density. With decreasing area, the laser power density or increasing the traverse speed, the melt pool volume decreases, as expected. Increasing the traverse speed also decreases the amount of powder deposited into the melt pool, further decreasing the melt pool volume. The profile of the deposited track was seen to be strongly dependent on the powder mass flow rate as it affects the thermal gradient in the melt pool. Possible columnar dendrites appear from the phase constrast at the rim of the melt pool at powder feedrate of 2 g min$^{-1}$, as marked in Figure 4a(ii), indicating excessive powder mass flow and the decrease of melt pool surface temperature. Further increasing the powder feedrate to 3 g $min^{-1}$, results in a distortion of the melt pool geometry as well as excessive un-melted powder on top of the melt track. Based on the melt pool geometry, 200W, 1 g $min^{-1}$ and 1 mm $s^{-1}$ were selected as the operating parameters for the diffraction experiment and the melt pool geometry is used to map out the diffreaction scanning strategies. This reduces undesirable features such as porosity and abnormal signals from the surface cooling and powders.

### 3.3 Phase transformations during cooling

Figure 5 shows the spatially resolved phase volume fractions, calculated from diffraction datasets, from the 1$^{st}$, 3$^{rd}$ and 5$^{th}$ layer of the thin wall built using the selected parameters. Figure 5a shows the average temperature in the X-ray beam direction in the mapped area, determined from lattice spacing expansion and contraction during laser heating and the subsequent cooling process. The melt pool temperature is set to be 1360 °C due to the limitations on detecting peak intensities.  As shown in Figure 5a, the temperature drops from the melting temperature to below 1200 °C within 1 mm of travel (1 s for the time duration) and the cooling rate decreases significantly within this regime. The average cooling rate in the mapped regime is ~150 to 500 K $s^{-1}$.

The molten liquid volume fraction, in the same regimes, was mapped to reveal the liquid-solid phase transformation, as shown in Figure 5b. Here, we assume that the integrated intensity of a peak associated with the liquid phase scales linearly with its volume fraction[36], interpolated between the intensities measured at 100% liquid and 100% solid. The limitation of this method is from the influence of complex chemical composition change over temperature to the integrated intensity, which will be further investigated. The thermal field was mapped with the liquid volume fraction evolution to understand the mushy zone and $\gamma$ phase development. Previously, such values could only be estimated using the dendrite arm spacing through metallographic examination[4,37]. The liquid volume fraction decreases to near zero at an indicated temperature of 1250°C, matching the reported solidus temperature (1260 °C) of IN718[37]. The region where a semi-solid phase composition is recorded is estimated to be around 500 μm along the build direction. The reasons for this semi-solid region are twofold; firstly, a mushy zone exists in the temperature range between the solidus and liquidus temperatures. Secondly, due to the 3-dimensional curvature of the melt-pool which is integrated through-thickness in the X-ray radiography and diffraction experiments. Therefore, this measurement represents an overestimation of the extent of the mushy zone.

The formation of the MC carbide phase and Laves phase during the solidification process is also mapped. The volume fraction of each phase is calculated in Figure 5d and 5e, respectively. The presence of texture in our study did not impact on the peak intesity. It is the first time the solidification sequence has been measured, *in situ,* along with the corresponding temperature and volume fraction information during DED-AM processing. The solidification process during the DED-AM melt track formation follows the sequence of the $\gamma$ phase, MC-type carbide and Laves phase, which conforms to prior modelling studies [38]. We observed that the solid phase emerged closer to the melt pool at the 3$^{rd}$ and the 5$^{th}$ layer and it is hypothesised that it is due to the changes of heat dissipation with a taller thin wall build.

### 3.4 Stress evolution

Diffraction data were collected in the melt track area, *in situ*, during the third layer of build, as shown in Figure 6. This data was used to quantify the stress evolution in contrast to the phase fraction during the DED-AM in a multi-layer build condition and can be used for cracking susceptibility prediction. $d_0$ in our study is the diffraction peaks of the substrate plate. The stresses measured in our experiments arise from a combination of thermal effects and solid-state phase transformation (volumetric effects)[39–41]. The temperature distribution of the region of interest is shown in Figure 6a and the liquid fraction of the same region is shown as Figure 6b. The out of plane stress, $\sigma_{zz}$, is not asscessible however, we assume it is negligible as the wall is thin. Furthermore, the in-plane stress is domimant due to the parallel direction to the laser, which shows highest thermal gradient. Thus, the in-plane stress components ($\sigma_{xx}$, $\sigma_{xy}$, and $\sigma_{yy}$) were calculated, as shown in Figure 6c. **The methodology to obtain these values can be found in the Supplementary Methods(2.2.1)**. The results show that the stress during the liquid-solid phase transformation is negligible (green region) and the stress starts to accumulate during solid-state phase transformation process (blue region), as shown in Figure 6b. $\sigma_{yy}$, parallel to the laser, appears to be the dominant stress component, fluctuating between $\pm$ 5 MPa before it increases gradually to 100 MPa as the temperature decreases to 1000 °C. The stress distribution in the three lines follows the temperature gradient. We hypothesise that $\sigma_{yy}$ is the dominant component due to the steep thermal gradient along the direction of the laser-induced rapid cooling. $\sigma_{xx}$, parallel to the build direction, decreases to – 70 MPa after a sudden increase to 35 MPa at 1000 °C during the cooling process in line1 and line2. In line3, $\sigma_{xx}$ gradually decreases to -100 MPa.

The post-build residual stress is also measured. The averaged internal residual stress of each build layer was calculated from diffraction patterns. The normal components of the residual stress are observed to be compression stress, ranging from -45 MPa to -125 MPa in the DED-AM build. The shear stress ($\sigma_{xy}$) component in the built sample is negligible, as a result, the difference between $\sigma_{xx}$ and $\sigma_{yy}$ is only ~ 5 to10 MPa. Thus, the resulting stress state is primarily hydrostatic compression.

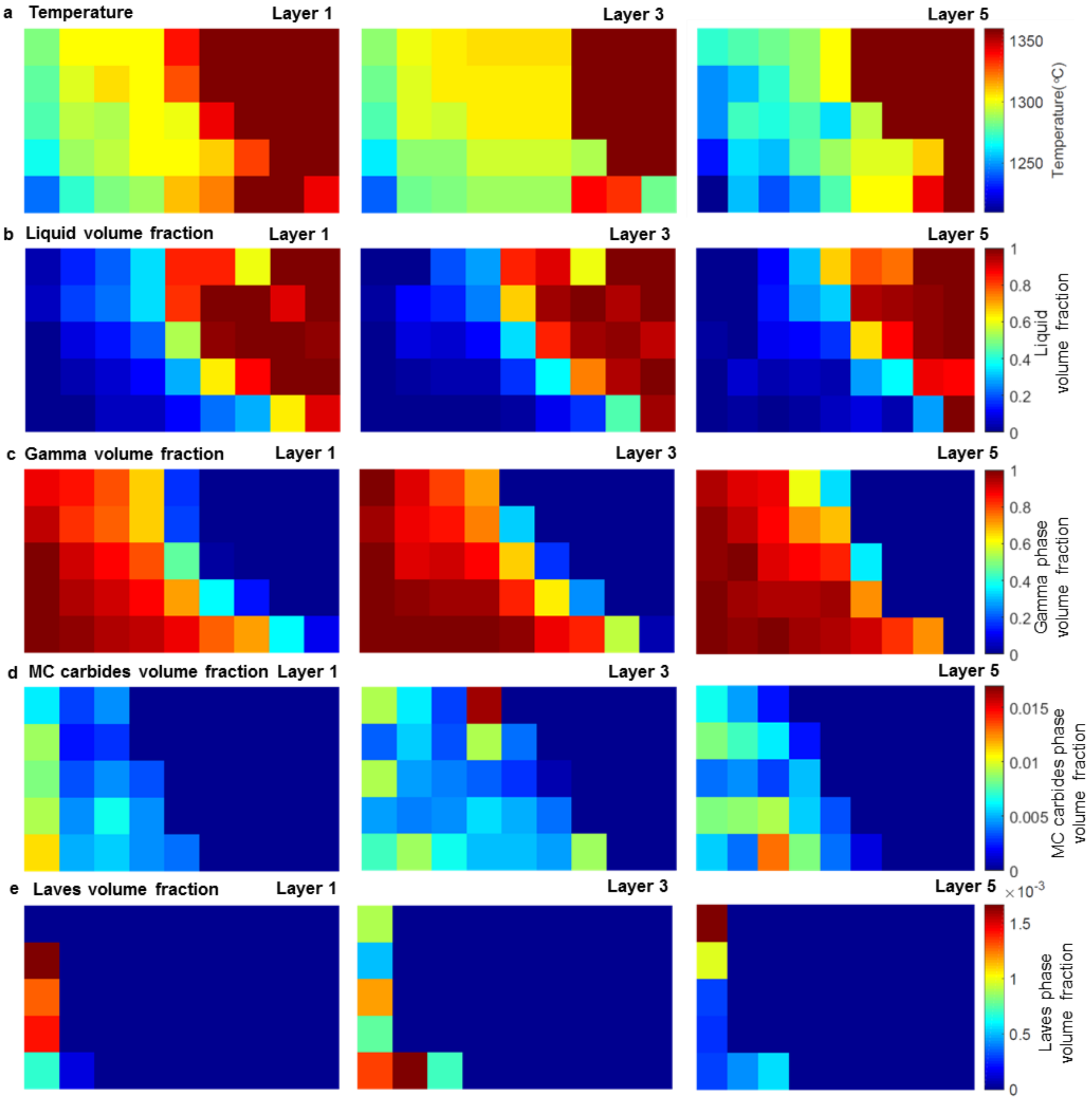

*Figure 5. Melt pool mapping results derived from diffraction patterns showing (a) Temperature, (b) liquid volume fraction, (c) γ phase volume fraction, (d) MC carbide volume fraction, (e) laves volume fraction in multi-layer melt pool region. The liquid volume fractions mapping results match the radiography of the melt pool boundary, the results can be found in Supplementary Information.*

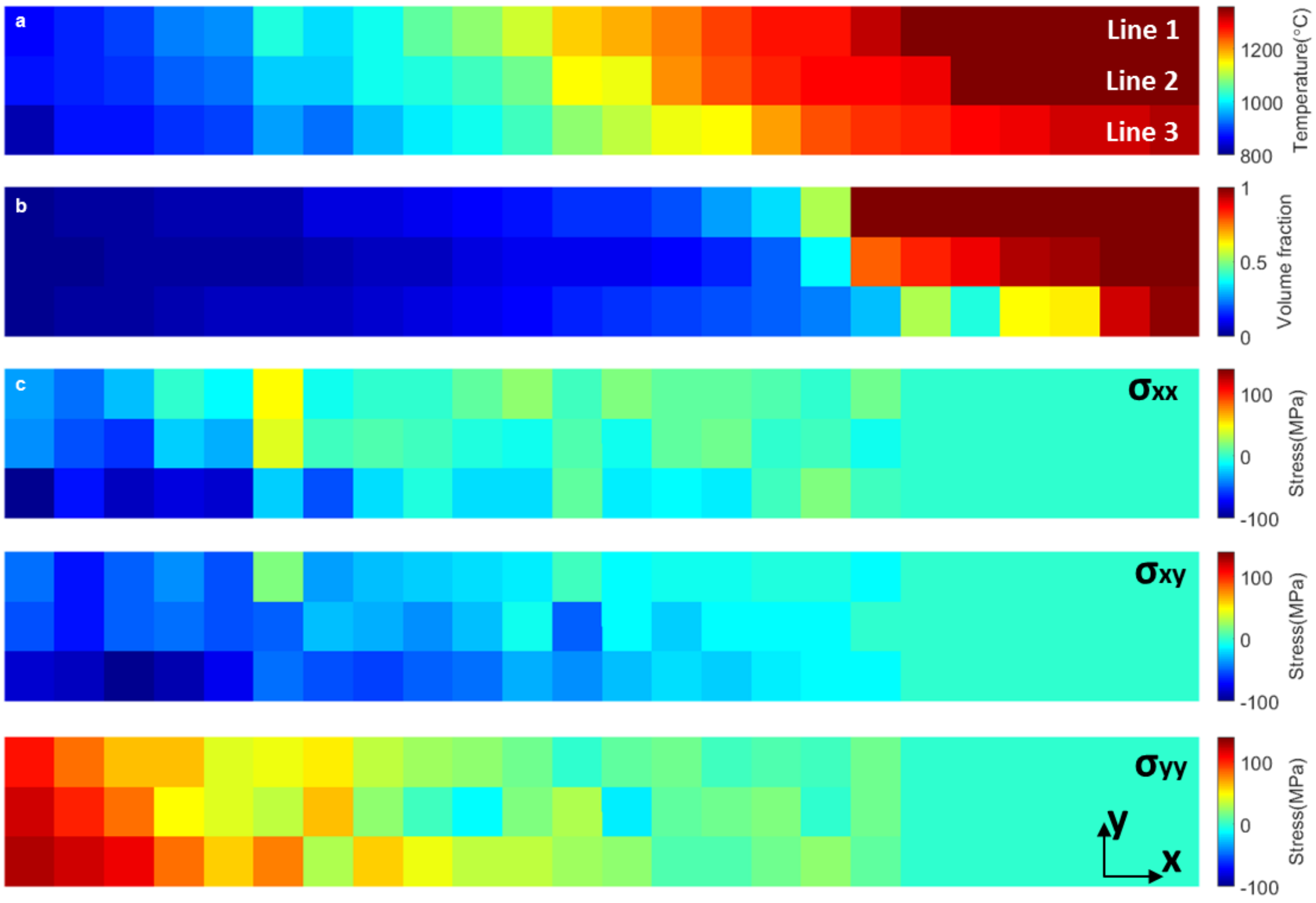

*Figure 6. Melt track mapping results derived from diffraction patterns. (a) Temperature distribution in multi-layer melt pool region including the newly formed melt track, the re-melt region between two layers and re-heat region (Heat Affected Zone, HAZ) of the previous layer during the third layer of build (b) Liquid phase volume fraction in the corresponding melt pool region. (c) Stress distribution including $\sigma_{xx}$, $\sigma_{yy}$, and $\sigma_{xy}$ in the melt track area. The errors of the stress calculations can be found in Supplementary Information.*

In this mapped region(Figure 6), the γ phase steadily increases to ~98.5 vol% at an average temperature of 800 °C through projection, a very small amount Laves phase (~0.05 vol%) and MC Carbides phase (~1.5 vol%) gradually develop in this region till the detected area. This is likely to be occurring close to the centreline of the deposit where the temperature remains significantly higher than that at the periphery. **Details of the volume fraction calculations can be found in Supplementary Methods(2.4)**. Although the evolution of secondary phases (MC carbides and Laves) are associated with transformation strains due to the volumetric mismatch and resultant deformation in the crystal structure, causing reconstructive stress [37], we hypothesise that as the volume fraction of the secondary phases is very small in the mapped region, the thermal effects are the dominant cause of stress.

# 4. Discussion

## 4.1 Marangoni flow

The solidification process in DED-AM is governed by the heat transfer in the melt pool[4], while the melt pool flow in DED-AM is a critical parameter which determines the heat and mass transfer during melt track development and microstructural feature formation. Tracing Marangoni flow during DED-AM process enables us to understand the melt pool evolution under constant laser radiation and powder incorporation. Although other factors such as the recoil pressure[10], buoyancy force[42], vapour pressure and shadowing effect[43] can also impact the melt flow, it is evident in our study that surface tension is the dominant driving force that governs the melt pool flow behaviour. We observed a radially outward flow in the melt pool and it produces a wide and shallow melt pool. Based on the Heiple–Roper theory[44], we confirm that the melt pool follows a negative surface tension gradient where the surface tension reduces with an increase of temperature. The melt flow indicates the speed of heat transfer and defines the temperature gradient based on Fourier's law. It is noted that our measured flow speed is lower than that indicated by prior modelling results[45]. This is attributed in part to our measurements being 2D radiographic projections of the 3D flow (Figure 4a), and hence they can be considered as the lower bounds of the true flow behaviour. As W particles are the same size as IN718 powder particles in this study, the size effect of the tracer is negligible. We speculate that the turbulence in the melt flow reduces the maximum velocity due to the temperature dependence of thermal diffusivity[46], and this has normally been neglected in the modelling calculations. In our study, to accurately describe the melt pool flow, we carefully choose tracer trajectories that appear to move in the plane orthogonal to the X-ray beam,flowing towards/from the extremities of the melt pool. **The instantaneous tracer velocity measurements can be found in Supplementary Results(3.1),** revealing the range of melt flow velocities in the melt pool.

Based on the quantitative analysis, the kinetic energy and heat transfer in the melt pool during DED-AM process can be interpreted using the Weber number ($W_e$) and the Péclet number (*Pe*). These values describe the laser melting process and are valuable input for LAM modelling process. The Weber number ($W_e$) is calculated to be 43 near the melt pool centre and 0.19 near the melt pool tail (**details can be found in Supplementary Methods(2.7)**). It shows that compared to LPBF[20], the difference of the Weber number between the melt pool centre and the melt pool tail is not significant, resulting in low surface roughness. To describe heat transfer in the melt pool, the Péclet number (*Pe*), which combines the Reynolds number ($Re_L$) and Prandtl number (*Pr*)[20], was used. The *Pe* number is calculated to be 107 near the melt pool centre and 1.3 at the solidification front. **The details of the calculation can be found in Supplementary Results(3.1.3).** The results indicate that heat convection dominates the centre of the melt pool while heat conduction dominates at the tail of the melt pool.

### 4.2 Solidification process and cooling rate

Solidification of the melt pool is governed by the net heat transfer through the melt pool[4] and in this study, we focus on two major events: (1) the mushy-zone heat transfer and (2) the microstructural evolution. The mushy-zone contains the solid, together with interdendritic liquid enriched in alloying elements. The heat transfer at the solid/liquid boundary has been indicated in section 4.1 by the $P_e$ number. Here, the thermal gradient is derived from the melt pool temperature mapping results using diffraction patterns, enabling a more accurate characterisation of melt pool solidification front behaviour. Previously, the thermal field could only be estimated by thermocouples, infrared thermography or modelling. Using $\gamma$ reflections that are seen to increase in intensity during solidification, the temperature gradient of the solidification front was calculated to be ~ 2.5 to 5 × $10^6$ K/mm, close to the predicted values via modelling[37], but different from welding and Laser Powder Bed Fusion process due to the powder input and scanning speed. Using this gradient, the lower velocity threshold for cellular-dendritic transition is calculated to be ~0.02 mm $s^{-1}$ (**the calculation method can be found** **in Supplementary Methods(2.5)**), well below the operating traverse speed (1 mm $s^{-1}$) in our experiments, promoting epitaxial dendritic growth.

### 4.3 Final solidification phase evolution

The rapid laser-induced heating and cooling rates during DED-AM has previously meant that the phases developed during DED-AM can only be measured by post build metallographic analysis[47], inferring the kinetics only through simulations[38]. Zhao *et al.*[48] indicated the possibility of the solidification rate estimation using X-ray imaging. However, our study directly quantifies the solidification sequence, including estimates of the phase formation temperatures, kinetics and volume fractions. We quantitatively determine that the major phases during DED-AM of IN718 are identified as γ, MC-type carbides and Laves for the process parameters used in this study.

We assume that during the cooling from the melt, the decrease of lattice spacing is solely due to thermal contraction[13]; therefore, the temperature evolution can be calculated from the lattice spacing using known thermal expansion coefficients. We also assume that the thermal contraction behaviour of the γ phase in IN718 is linear from the melt to room temperature[49]. The temperature evolution is then calculated from the lattice spacing using thermal expansion coefficients tabulated from the liquidus and room temperature. Other than the formation of MC carbides and the Laves phase, the IN718 samples do not exhibit further phase transformations during cooling in the mapped region as the rapid cooling rate is considered to be sufficient to suppress the formation of γ' phase and γ'' phase. Without any further solid-state transformations, it is possible to separate the contribution of thermal effects from chemical and stress effects.

Based on our observations, the solidification path starts with the formation of epitaxial dendritic γ at 1275 -1300 °C. Secondary phases start to form in the interdendritic liquid when the solute elements (i.e. Nb, Ti and C) are likely to exceed their solubility limit. For IN718, the enrichment of Nb and Ti in the interdendritic liquid eventually leads to the formation of MC-type carbides (TiC) via a non-invariant eutectic type reaction. The temperature of this reaction in our experiments is observed to be 1250 °C in the first layer and 1250-1275 °C in the subsequent layers, which agrees well with the calculation performed by *Knorovsky et al.* [38] (exactly at 1250 °C). We suggest that the difference in the reaction temperature in the subsequent layers is due to the subtle changes in alloy composition from the remelting of previous layers. The formation of carbides depletes the carbon in the interdendritic liquid, enriching the remaining elements that advance the $\gamma$/Laves eutectic reaction at 1200 °C. Once carbides and Laves form, their volume fractions continue to increase (marked in Figure 6b). It is difficult to estimate their growth rate due to their low volume fraction. ~1.1 vol% of MC carbides and ~0.5 vol% of Laves are detected post-mortem using diffraction in this thin wall build, which is similar to the values reported in an industrial DED-AM process. Sui *et al*.[50] reported 1.55 vol% Laves phase using a LSF-IIIB LAM system at room temperature. The differences observed can be explained by the differences in process conditions particularly cooling rate. In contrast to the reported complex constituents of (α-Cr + δ + Laves + $\gamma''$ + matrix) [47,51], no traces of δ and $\gamma''$ were detected in our study. We hypothesise that the high-temperature gradient and rapid solidification rate inhibits time-dependent diffusional processes including the formation of strengthening phases during DED-AM. Furthermore, no δ was detected in the post-mortem build, and we further hypothesise that it is because the δ phase nucleation kinetics are too slow for the cooling rate [47] imposed by DED-AM in our study.

### 4.4 Cracking criteria

Laser-induced rapid heating and cooling produce a steep thermal gradient, resulting in significant volumetric shrinkage and residual stresses. Meanwhile, the eutectic reaction and elemental segregation at the grain boundaries and interdendritic regions generated significant undercooling[52] from the solute enrichment during LAM of IN718. The combination of intergranular liquid thin films and high thermal stresses increases the susceptibility for hot cracking, especially liquation cracking in the Heat Affected Zone[26]. Fundamentally, two conditions must be satisfied for hot cracking to occur: (a) mechanically/thermally imposed restrain (strain) exhausts the ductility of the material; and (b) a crack-susceptible microstructure results from the persistence of liquid films along solidification boundaries[53]. To understand the cracking susceptibility of IN718 in DED-AM, the von Mises stresses in the melt track region were first calculated to estimate the yielding of IN718 during the rapid cooling process. The von Mises stresses were calculated for measured locations in the melt track calculated from the stress tensor components evaluated in section 3.4. The values are plotted as a function of temperature in Figure 7a. To predict the yielding of IN718, yield strengths[54] at a strain rate of 0.001 $s^{-1}$ are also plotted in Figure 7a. **Calculation of the appropriate strain rate in the region of interest can be found in the Supplementary Methods(2.3).**

Figure 7a shows that all stresses are zero at the liquidus temperature (1360 °C) in the region of interest and they increase during cooling. The von Mises stress of line 1 and 2 are close to the reported yield strength of IN718, while the re-melting of the previous build layer causes the stress increase. This results in the measured von Mises stress of line 3 being approximately equal to the yield strength. Therefore, during rapid cooling from the DED-AM of IN718, the material reaches the yield stress immediately once solid forms. It can be inferred that as the material reaches near 100% solid, plastically must accumulate which prevails continuously as the temperature decreases. However, as no $\gamma$' precipitates nor other strengthening phases are detected, any heterogeneous nucleation mechanisms remain insufficient to overcome the kinetically inhibiting rapid cooling rate, as discussed in 4.3. Whilst significant plasticity is generated at high temperature, there was no evidence of recrystallisation, which we postulate is due to the same reasons as the lack of precipitation.

The solidification microstructure is another criterion for cracking-susceptibility. Many hot cracking models, e.g. the Rappaz, Drezet and Gremaud (RDG) [55], indicate that cracking is dominated by the final stage of the solidification process. The susceptibility for an alloy to crack is often interpreted from Scheil-Gulliver solidification models using the fraction of the solid. Alloys known to crack exhibit a large solidification range (the difference between the liquidus and solidus temperature) and have a sharp turnover in the solidification curves at high fractions of solid[56]. In our study, Thermo-Calc software was used to calculate the fraction of solid and the composition of solid-state phases during cooling, for phases known to form from the diffraction patterns. **Details of the simulation can be found in the Supplementary Methods(2.4.2).** Figure 7b shows the fraction of solid decreasing during the cooling process and following the simulated curve in the region of interest during the cooling process. A high solidification range and steep solid fraction gradients at high fractions of solids (greater than 0.8) were observed from the experimental results. Combining the two cracking criteria, we conclude that IN718 is highly susceptible to hot cracking during DED-AM. This conforms with the findings of Chauvet *et al.*[57] and Chen *et al.*[26]. Considering that plasticity prevails from re-melting, liquation cracking is most likely to occur in our study from the localised melting of the grain boundaries. Lippold *et al*.[53] explained that it is due to the addition of Ti or Nb, which forms MC-type carbides, in IN718. The cracking phenomenon is evidenced by the release of shear stress in the post-build residual stress measurements. Due to the intergranular nature of the cracks, the normal components of the residual stress are still detectable in the post-build scans, which can be further released by heat-treatment.

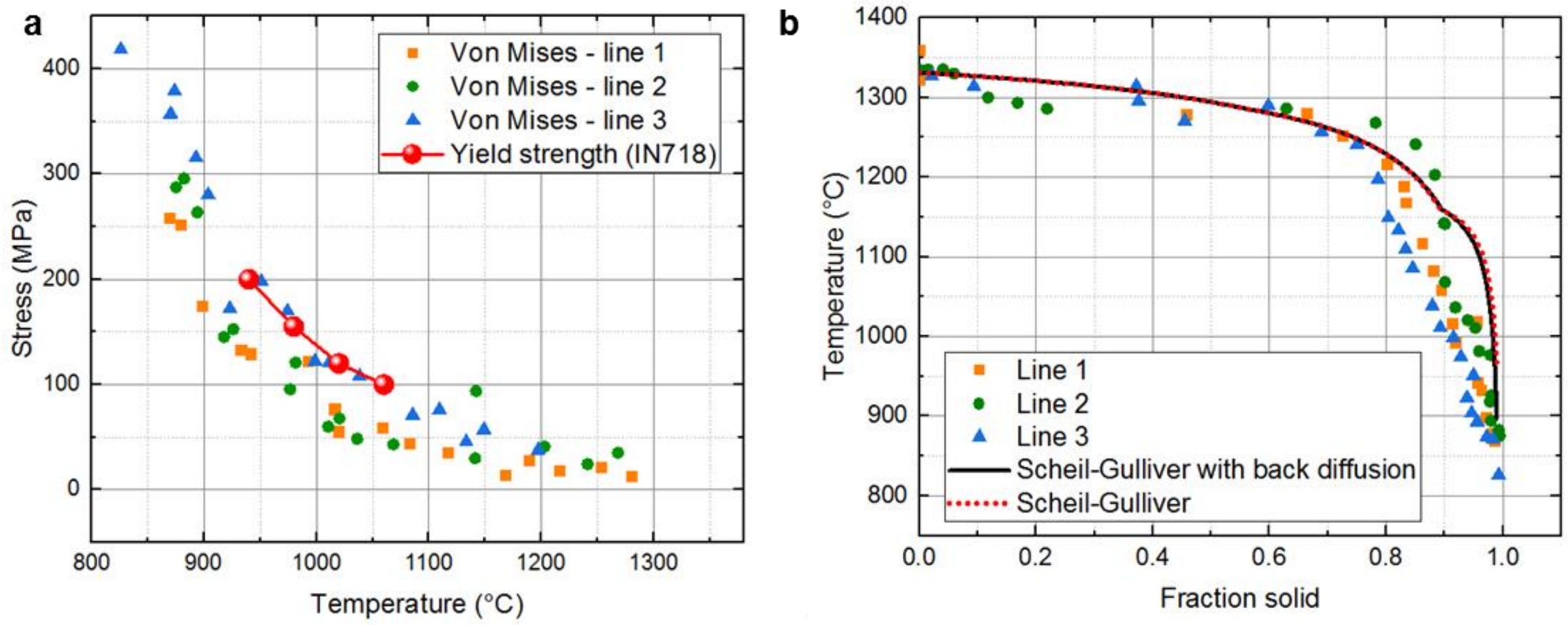

*Figure 7. Cracking susceptibility of IN718 during DED-AM. (a) Von Mises stresses in the melt track region calculated against IN718 yield strength at a strain rate of 0.001 s$^{-1}$. Plasticity accumulates which prevails continuously as the temperature decreases.* ***For error calculation see Supplementary Methods(2.2.2).*** *(b) The fraction of solids during the cooling process showing a large solidification range.*

## 5. Conclusions

A unique *in situ* and *operando* process replicator has been successfully implemented that enabled fast time-resolved synchrotron real and reciprocal space X-ray imaging of the underlying physics occurring during multi-layer DED-AM. Significant new insights were obtained that relate the laser melting to the resultant microstructure and potentially detrimental features such as porosity and micro-cracks during the DED-AM of the nickel-based superalloy, IN718. The following key conclusions can be drawn:

- Using imaging to guide the analysis of temporally resolved diffraction data, it was possible to unambiguously separate the thermal gradient, phase transformations including their volume fraction development and macromechanical stress development during a multi-layer DED-AM build.
- By measuring the instantaneous cooling rate, it can be hypothesised that the solidification front in IN718 during DED-AM is dendritic. During cooling, the following phase transformation sequence is observed: Liquid → Liquid + $\gamma$ → Liquid + $\gamma$ + MC → Liquid + $\gamma$ + MC + Laves → $\gamma$ + MC + Laves. The as-built microstructure comprises mostly $\gamma$, but also contains low volume fraction phases including; MC carbides (~1.1 vol.%) Laves (~0.5 vol.%), demonstrating the high sensitivity of the technique.
- The heat transfer kinetics in the melt pool were quantified by tracing the speed and directions of the melt flow. Marangoni convection is the prevailing phenomenon that governs the melt pool flow, whereby heat convection controls the centre of the melt pool while in the mushy zone, heat conduction dominates.
- It was revealed that the accumulated stresses in the high-temperature region are dominated by thermal effects. The stress is sufficiently high to yield the material immediately after it solidifies and during further cooling. Instead of triggering the formation of secondary strengthening phases or recrystallization, the stresses exhaust the ductility of the material and likely lead to liquation cracking due to the large solidification range.

Through characterising the transient behaviour operative during the DED-AM process, it is possible to understand the as-built characteristics and can be used to design post-build heat treatment strategies. The application of such insight, both to nickel-base superalloys and other alloy systems, will deliver critical understanding in enhancing the quality and performance of additively manufactured parts.

**Acknowledgements**

This research was supported under MAPP: EPSRC Future Manufacturing Hub in Manufacture using Advanced Powder Processes (EP/P006566/1), a Royal Academy of Engineering Chair in Emerging Technology (CiET1819/10), and Rolls-Royce plc. via the Horizon 2020 Clean Sky 2 WP5.8.1 programme. This work has been supported by the Office of Naval Research (ONR) Grant N62909-19-1-2109. The provision of materials and technical support from Rolls-Royce plc is gratefully acknowledged. Laboratory space and facilities were provided by the Research Complex at Harwell. The authors thank Diamond Light Source for providing beamtime (MT20096) and the staff at I12 beamline for technical assistance.

# References

[1] S.A.M. Tofail, E.P. Koumoulos, A. Bandyopadhyay, S. Bose, L. O'Donoghue, C. Charitidis, Additive manufacturing: scientific and technological challenges, market uptake and opportunities, Mater. Today. 21 (2018) 22–37. doi:10.1016/J.MATTOD.2017.07.001.

[2] Z. Quan, A. Wu, M. Keefe, X. Qin, J. Yu, J. Suhr, J.H. Byun, B.S. Kim, T.W. Chou, Additive manufacturing of multi-directional preforms for composites: Opportunities and challenges, Mater. Today. 18 (2015) 503–512. doi:10.1016/j.mattod.2015.05.001.

[3] T.B. Sercombe, G.B. Schaffer, Rapid manufacturing of aluminum components, Science (80-. ). 301 (2003) 1225–1227. doi:10.1126/science.1086989.

[4] S.M. Thompson, L. Bian, N. Shamsaei, A. Yadollahi, An overview of Direct Laser Deposition for additive manufacturing; Part I: Transport phenomena, modeling and diagnostics, Addit. Manuf. 8 (2015) 36–62. doi:10.1016/J.ADDMA.2015.07.001.

[5] T.M. Pollock, Alloy design for aircraft engines, Nat. Mater. 15 (2016) 809–815. doi:10.1038/nmat4709.

[6] F. Trevisan, F. Calignano, A. Aversa, G. Marchese, M. Lombardi, S. Biamino, D. Ugues, D. Manfredi, Additive manufacturing of titanium alloys in the biomedical field: processes, properties and applications, J. Appl. Biomater. Funct. Mater. 16 (2018) 57–67. doi:10.5301/jabfm.5000371.

[7] A. Saboori, A. Aversa, G. Marchese, S. Biamino, M. Lombardi, P. Fino, Application of directed energy deposition-based additive manufacturing in repair, Appl. Sci. 9 (2019). doi:10.3390/app9163316.

[8] M. Megahed, H.-W. Mindt, N. N'Dri, H. Duan, O. Desmaison, Metal additive-manufacturing process and residual stress modeling, Integr. Mater. Manuf. Innov. 5 (2016) 61–93. doi:10.1186/s40192-016-0047-2.

[9] N. Shamsaei, A. Yadollahi, L. Bian, S.M. Thompson, An overview of Direct Laser Deposition for additive manufacturing; Part II: Mechanical behavior, process parameter optimization and control, Addit. Manuf. 8 (2015) 12–35. doi:10.1016/j.addma.2015.07.002.

[10] Y. Chen, S.J. Clark, C. Lun, A. Leung, L. Sinclair, S. Marussi, M.P. Olbinado, E. Boller, A. Rack, I. Todd, P.D. Lee, In-situ Synchrotron imaging of keyhole mode multi-layer laser powder bed fusion additive manufacturing, Appl. Mater. Today. 20 (2020) 100650. doi:10.1016/j.apmt.2020.100650.

[11] C.L.A. Leung, S. Marussi, R.C. Atwood, M. Towrie, P.J. Withers, P.D. Lee, In situ X-ray imaging of defect and molten pool dynamics in laser additive manufacturing, Nat. Commun. 9 (2018) 1–9. doi:10.1038/s41467-018-03734-7.

[12] S.M.H. Hojjatzadeh, N.D. Parab, W. Yan, Q. Guo, L. Xiong, C. Zhao, M. Qu, L.I. Escano, X. Xiao, K. Fezzaa, W. Everhart, T. Sun, L. Chen, Pore elimination mechanisms during 3D printing of metals, Nat. Commun. 10 (2019) 1–8. doi:10.1038/s41467-019-10973-9.

[13] S. Hocine, H. Van Swygenhoven, S. Van Petegem, C.S.T. Chang, T. Maimaitiyili, G. Tinti, D. Ferreira Sanchez, D. Grolimund, N. Casati, Operando X-ray diffraction during laser 3D printing, Mater. Today. 34 (2020) 30–40. doi:10.1016/j.mattod.2019.10.001.

[14] C. Zhao, K. Fezzaa, R.W. Cunningham, H. Wen, F. De Carlo, L. Chen, A.D. Rollett, T. Sun, Real-time monitoring of laser powder bed fusion process using high-speed X-ray imaging and diffraction, Sci. Rep. 7 (2017) 3602. doi:10.1038/s41598-017-03761-2.

[15] N.P. Calta, J. Wang, A.M. Kiss, A.A. Martin, P.J. Depond, G.M. Guss, V. Thampy, A.Y. Fong, J.N. Weker, K.H. Stone, C.J. Tassone, M.J. Kramer, M.F. Toney, A. Van Buuren, M.J. Matthews, An instrument for *in situ* time-resolved X-ray imaging and diffraction of laser powder bed fusion additive manufacturing processes, Rev. Sci. Instrum. 89 (2018) 055101. doi:10.1063/1.5017236.

[16] N. Kouraytem, X. Li, R. Cunningham, C. Zhao, N. Parab, T. Sun, A.D. Rollett, A.D. Spear, W. Tan, Effect of Laser-Matter Interaction on Molten Pool Flow and Keyhole Dynamics, Phys. Rev. Appl. 11 (2019) 1. doi:10.1103/PhysRevApplied.11.064054.

[17] Q. Guo, C. Zhao, L.I. Escano, Z. Young, L. Xiong, K. Fezzaa, W. Everhart, B. Brown, T. Sun, L. Chen, Transient dynamics of powder spattering in laser powder bed fusion additive manufacturing process revealed by in-situ high-speed high-energy x-ray imaging, Acta Mater. 151 (2018) 169–180. doi:10.1016/j.actamat.2018.03.036.

[18] R. Cunningham, C. Zhao, N. Parab, C. Kantzos, J. Pauza, K. Fezzaa, T. Sun, A.D. Rollett, Keyhole threshold and morphology in laser melting revealed by ultrahigh-speed x-ray imaging, Science (80-. ). 363 (2019) 849–852. doi:10.1126/science.aav4687.

[19] C.L.A. Leung, S. Marussi, M. Towrie, J. del Val Garcia, R.C. Atwood, A.J. Bodey, J.R. Jones, P.J. Withers, P.D. Lee, Laser-matter interactions in additive manufacturing of stainless steel SS316L and 13-93 bioactive glass revealed by in situ X-ray imaging, Addit. Manuf. 24 (2018) 647–657. doi:10.1016/J.ADDMA.2018.08.025.

[20] Q. Guo, C. Zhao, M. Qu, L. Xiong, S.M.H. Hojjatzadeh, L.I. Escano, N.D. Parab, K. Fezzaa, T. Sun, L. Chen, In-situ full-field mapping of melt flow dynamics in laser metal additive manufacturing, Addit. Manuf. 31 (2020) 100939. doi:10.1016/j.addma.2019.100939.

[21] M. Drakopoulos, T. Connolley, C. Reinhard, R. Atwood, O. Magdysyuk, N. Vo, M. Hart, L. Connor, B. Humphreys, G. Howell, S. Davies, T. Hill, G. Wilkin, U. Pedersen, A. Foster, N. De Maio, M. Basham, F. Yuan, K. Wanelik, I12: the Joint Engineering, Environment and Processing (JEEP) beamline at Diamond Light Source, J. Synchrotron Radiat. 22 (2015) 828–838. doi:10.1107/S1600577515003513.

[22] V. Thampy, A.Y. Fong, N.P. Calta, J. Wang, A.A. Martin, P.J. Depond, A.M. Kiss, G. Guss, Q. Xing, R.T. Ott, A. van Buuren, M.F. Toney, J.N. Weker, M.J. Kramer, M.J. Matthews, C.J. Tassone, K.H. Stone, Subsurface Cooling Rates and Microstructural Response during Laser Based Metal Additive Manufacturing, Sci. Rep. 10 (2020) 1–9. doi:10.1038/s41598-020-58598-z.

[23] S. Hocine, S. Van Petegem, U. Frommherz, G. Tinti, N. Casati, D. Grolimund, H. Van Swygenhoven, A miniaturized selective laser melting device for operando X-ray diffraction studies, Addit. Manuf. 34 (2020) 101194. doi:10.1016/j.addma.2020.101194.

[24] N.J. E., Improving turbine blade performance by solidification control, Metallurgia. 46 (1979) 437.

[25] T. Özel, D. Ulutan, Prediction of machining induced residual stresses in turning of titanium and nickel based alloys with experiments and finite element simulations, CIRP Ann. - Manuf. Technol. 61 (2012) 547–550. doi:10.1016/j.cirp.2012.03.100.

[26] Y. Chen, F. Lu, K. Zhang, P. Nie, S.R. Elmi Hosseini, K. Feng, Z. Li, Dendritic microstructure and hot cracking of laser additive manufactured Inconel 718 under improved base cooling, J. Alloys Compd. 670 (2016) 312–321. doi:10.1016/j.jallcom.2016.01.250.

[27] B.P. L., The mechanical and microstructural characteristics of laser-deposited IN718, J. Mater. Process. Technol. 170 (2005) 240.

[28] F. G., Microstructure and mechanical properties of nickel based superalloy IN718 produced by rapid prototyping with electron beam melting (EBM), Mater. Sci. Technol. 27 (2011) 876.

[29] Z. X., The microstructure and mechanical properties of deposited-IN718 by selective laser melting, J. Alloy. Compd. 513 (2012) 518.

[30] A.M. M., Microstructural and texture development in direct laser fabricated IN718, Mater. Charact. 89 (2014) 102.

[31] D. Zhang, W. Niu, X. Cao, Z. Liu, Effect of standard heat treatment on the microstructure and mechanical properties of selective laser melting manufactured Inconel 718 superalloy, Mater. Sci. Eng. A. 644 (2015) 32–40. doi:10.1016/j.msea.2015.06.021.

[32] Y. Chen, S.J. Clark, L. Sinclair, C.L.A. Leung, S. Marussi, T. Connolley, O. V. Magdysyuk, R.C. Atwood, G.J. Baxter, M.A. Jones, D.G. McCartney, I. Todd, P.D. Lee, In situ and Operando X-ray Imaging of Directed Energy Deposition Additive Manufacturing, ArXiv. (2020) 1–31. http://arxiv.org/abs/2006.09087.

[33] J.-Y. Tinevez, N. Perry, J. Schindelin, G.M. Hoopes, G.D. Reynolds, E. Laplantine, S.Y. Bednarek, S.L. Shorte, K.W. Eliceiri, TrackMate: An open and extensible platform for single-particle tracking, Methods. 115 (2017) 80–90. doi:https://doi.org/10.1016/j.ymeth.2016.09.016.

[34] J. Schindelin, I. Arganda-Carreras, E. Frise, V. Kaynig, M. Longair, T. Pietzsch, S. Preibisch, C. Rueden, S. Saalfeld, B. Schmid, J.-Y. Tinevez, D.J. White, V. Hartenstein, K. Eliceiri, P. Tomancak, A. Cardona, Fiji: an open-source platform for biological-image analysis, Nat. Methods. 9 (2012) 676–682. doi:10.1038/nmeth.2019.

[35] J. Filik, A.W. Ashton, P.C.Y. Chang, P.A. Chater, S.J. Day, M. Drakopoulos, M.W. Gerring, M.L. Hart, O. V. Magdysyuk, S. Michalik, A. Smith, C.C. Tang, N.J. Terrill, M.T. Wharmby, H. Wilhelm, Processing two-dimensional X-ray diffraction and small-angle scattering data in DAWN 2, J. Appl. Crystallogr. 50 (2017) 959–966. doi:10.1107/S1600576717004708.

[36] H.P. Wang, C.H. Zheng, P.F. Zou, S.J. Yang, L. Hu, B. Wei, Density determination and simulation of Inconel 718 alloy at normal and metastable liquid states, J. Mater. Sci. Technol. 34 (2018) 436–439. doi:10.1016/j.jmst.2017.10.014.

[37] T. DebRoy, H.L. Wei, J.S. Zuback, T. Mukherjee, J.W. Elmer, J.O. Milewski, A.M. Beese, A. Wilson-Heid, A. De, W. Zhang, Additive manufacturing of metallic components – Process, structure and properties, Prog. Mater. Sci. 92 (2018) 112–224. doi:10.1016/j.pmatsci.2017.10.001.

[38] G.A. Knorovsky, M.J. Cieslak, T.J. Headley, A.D. Romig, W.F. Hammetter, INCONEL 718: A

solidification diagram, Metall. Trans. A. 20 (1989) 2149–2158. doi:10.1007/BF02650300.

[39] L.E. Lindgren, Finite element modeling and simulation of welding part 1: Increased complexity, J. Therm. Stress. 24 (2001) 141–192. doi:10.1080/01495730150500442.

[40] L.E. Lindgren, Finite element modeling and simulation of welding. part 2: Improved material modeling, J. Therm. Stress. 24 (2001) 195–231. doi:10.1080/014957301300006380.

[41] L.E. Lindgren, Finite element modeling and simulation of welding. Part 3: Efficiency and integration, J. Therm. Stress. 24 (2001) 305–334. doi:10.1080/01495730151078117.

[42] S.J. Wolff, H. Wu, N. Parab, C. Zhao, K.F. Ehmann, T. Sun, J. Cao, In-situ high-speed X-ray imaging of piezo-driven directed energy deposition additive manufacturing, Sci. Rep. 9 (2019) 1–14. doi:10.1038/s41598-018-36678-5.

[43] F. Lia, J. Park, J. Tressler, R. Martukanitz, Partitioning of laser energy during directed energy deposition, Addit. Manuf. 18 (2017) 31–39. doi:10.1016/j.addma.2017.08.012.

[44] C.R. Heiple, J.R. Rope, R.T. Stagner, R.J. Aden, Surface Active Element Effects on the Shape of Gta, Laser, and Electron Beam Welds., Weld. J. (Miami, Fla). 62 (1983) 72–77.

[45] S. Wen, Y.C. Shin, Modeling of transport phenomena during the coaxial laser direct deposition process, J. Appl. Phys. 108 (2010). doi:10.1063/1.3474655.

[46] N. Chakraborty, S. Chakraborty, Distinct influences of turbulence in momentum, heat and mass transfers during melt pool convection in a typical laser surface alloying process, Eur. Phys. J. Appl. Phys. 36 (2006) 71–89. doi:10.1051/epjap.

[47] R.G. Ding, Z.W. Huang, H.Y. Li, I. Mitchell, G. Baxter, P. Bowen, Electron microscopy study of direct laser deposited IN718, Mater. Charact. 106 (2015) 324–337. doi:10.1016/j.matchar.2015.06.017.

[48] C. Zhao, K. Fezzaa, R.W. Cunningham, H. Wen, F. De Carlo, L. Chen, A.D. Rollett, T. Sun, Real-time monitoring of laser powder bed fusion process using high-speed X-ray imaging and diffraction, Sci. Rep. 7 (2017) 1–11. doi:10.1038/s41598-017-03761-2.

[49] D. Ramachandran, A.M. Kamalan Kirubaharan, A.M. Rabel, P. Kuppusami, Thermal expansion behaviour of inconel-690 by in-situ high temperature X-ray diffraction, Mater. Sci.

Forum. 830–831 (2015) 367–370. doi:10.4028/www.scientific.net/MSF.830-831.367.

[50] S. Sui, H. Tan, J. Chen, C. Zhong, Z. Li, W. Fan, A. Gasser, W. Huang, The influence of Laves phases on the room temperature tensile properties of Inconel 718 fabricated by powder feeding laser additive manufacturing, Acta Mater. 164 (2019) 413–427. doi:10.1016/j.actamat.2018.10.032.

[51] Y.L. Kuo, S. Horikawa, K. Kakehi, The effect of interdendritic δ phase on the mechanical properties of Alloy 718 built up by additive manufacturing, Mater. Des. 116 (2017) 411–418. doi:10.1016/j.matdes.2016.12.026.

[52] W. Kurz, Solidification microstructure-processing maps: Theory and application, Adv. Eng. Mater. 3 (2001) 443–452. doi:10.1002/1527-2648(200107)3:7<443::AID-ADEM443>3.0.CO;2-W.

[53] J.C. Lippold, Welding Metallurgy and Weldability, 2014. doi:10.1002/9781118960332.

[54] F. Chen, J. Liu, H. Ou, B. Lu, Z. Cui, H. Long, Flow characteristics and intrinsic workability of IN718 superalloy, Mater. Sci. Eng. A. 642 (2015) 279–287. doi:10.1016/j.msea.2015.06.093.

[55] M. Rappaz, J.M. Drezet, M. Gremaud, A new hot-tearing criterion, Metall. Mater. Trans. A Phys. Metall. Mater. Sci. 30 (1999) 449–455. doi:10.1007/s11661-999-0334-z.

[56] J.H. Martin, B.D. Yahata, J.M. Hundley, J.A. Mayer, T.A. Schaedler, T.M. Pollock, 3D printing of high-strength aluminium alloys, Nature. 549 (2017) 365–369. doi:10.1038/nature23894.

[57] E. Chauvet, P. Kontis, E.A. Jägle, B. Gault, D. Raabe, C. Tassin, J.J. Blandin, R. Dendievel, B. Vayre, S. Abed, G. Martin, Hot cracking mechanism affecting a non-weldable Ni-based superalloy produced by selective electron Beam Melting, Acta Mater. 142 (2018) 82–94. doi:10.1016/j.actamat.2017.09.047.

# *In situ* X-ray Investigation of Directed Energy Deposition Additive Manufacturing

Yunhui Chen[1,2*], Samuel J. Clark[1,2], David M. Collins[3], Sebastian Marussi[1,2], Simon A. Hunt[4], Danielle M. Fenech[5],Thomas Connolley[6], Robert Atwood[6], Oxana V. Magdysyuk[6], Gavin Baxter[7], Martyn Jones[7], Chu Lun Alex Leung[1,2], Peter D. Lee[1,2*]

[1]Mechanical Engineering, University College London, Torrington Place, London WC1E 7JE, UK
[2]Research Complex at Harwell, Rutherford Appleton Laboratory, Oxfordshire OX11 0FA, UK
[3]School of Metallurgy and Materials, University of Birmingham, Edgbaston, Birmingham B15 2TT, UK
[4]Department of Materials, University of Manchester, Oxford Rd, Manchester M13 9PL, UK
[5]Department of Physics, Cavendish Laboratory, University of Cambridge, JJ Thompson Avenue,Cambridge, Cambridgeshire CB3 0HE, UK
[6]Diamond Light Source, Harwell Campus, Oxfordshire, OX11 0DE, UK
[7]Rolls-Royce plc, PO Box 31, Derby, DE24 8BJ, UK

*Correspondence authors: peter.lee@ucl.ac.uk, yunhui.chen@ucl.ac.uk

# 1 Supplementary Information

## 1.1 The Blown Powder Additive Manufacturing Process Replicator (BAMPR)

The system was encased within a Class I laser enclosure and comprises an inert environment chamber, a high precision 3-axis platform (Aerotech, US), a coaxial DED nozzle and a laser system (SPI Lasers Ltd, UK). An industrial powder feeder (Oerlikon Metco TWIN-10-C) delivers powder to the system in a stream of argon gas. The laser is a 1070 nm Ytterbium-doped fibre laser (continuous-wave mode) with controllable laser power ($P$) in the range of 0-200 W. The laser is coupled with tuneable optics (Optogama, Lithuania) to facilitate a controllable focused spot size of between 200-700 µm. This allows the BAMPR to operate at power density of 397 - 6366 W mm$^{-2}$, replicating the range of an industrial DED AM machine. The laser is positioned to be concentric with the powder delivery stream blown from the nozzle and normal to the substrate plate under an Argon carrier gas. In this work the build platform was able to translate through 50 mm in width, 25 mm in length and 50 mm in height. The substrate plate is positioned inside the environmental build chamber with Kapton windows for X-ray transparency. The chamber is constantly refreshed with a combined flowing argon atmosphere of 17 L min$^{-1}$. The speed of the sample stages in both cases were controlled to be between 1 - 5 mm s$^{-1}$ to enable a continuous track to be formed, as proved successful in preliminary laboratory trials.

## 1.2 Materials

The morphology of the Plasma Rotate Electrode Process (PREP) IN718 powder was characterized by a JEOL JSM-6610LV SEM. The particle size distribution was measured from SEM micrographs which were analysed using the Image Processing Toolbox in MATLAB 2016a (The MathWorks Inc, USA). The chemical states of the virgin powder were examined by ICP-OES.

## 1.3 Synchrotron Imaging and Diffraction Acquisition Conditions

We performed *in situ* X-ray imaging on the I12-JEEP beamline at the Diamond Light Source to capture the transient phenomena of DED-AM in IN718. A monochromatic beam was selected to provide the best signal to noise level, and ensure linear attenuation. A mean X-ray energy of approximately 53 keV was used for all experiments. The X-ray imaging system consisted of a 200 µm thick LuAg:Ce scintillator and a 4× magnification using a long working distance objective lens (0.21 numerical aperture). The X-rays were attenuated by the powder and the deposit which were converted to visible light by a scintillator, lens-coupled to a high-resolution imaging camera (PCO.edge, PCO). This recorded images at 200 fps and a lower resolution high-speed camera (MIRO 310M, Vision Research Inc.), recording at up to 5,000 fps. Unlike for laser powder bed fusion, the melt pool in DED-AM is much larger and the translations are much slower (mm $s^{-1}$ compared to m $s^{-1}$). Therefore, we optimised the number of frames per second for two situations. Firstly, to capture the pool phenomena we found 200 fps was sufficient to measure the key phenomena, while enabling a long exposure time of 0.0049 s and a very small off-load time (0.0001 s) provided an excellent signal to noise (S/N) ratio. We also used 5000 fps for experiments where faster phenomena were occurring, trading off S/N for speed. The optical configuration provided an imaging resolution of approximately 6.67 µm per pixel at 5,000 fps and 3.24 µm per pixel at 200 fps.

We performed *in situ* X-ray diffraction on the I12-JEEP beamline at Diamond Light Source to capture the transient phenomena of the DED-AM of IN718. A monochromatic beam with mean X-ray energy of approximately 70 $keV$ was used and calibrated using a $CeO_2$ standard for all diffraction experiments. *In situ* diffraction data were acquired using a large Pilatus 2M CdTe 2D area detector, with an active area of 253.7 × 288.8 $mm^2$ (1475 × 679 $pixels^2$) and pixel size of 172 × 172 $µm^2$, enabling full Debye-Scherrer diffraction patterns, obtained in transmission, to be collected with an exposure time of 6.67 s.

# 2 Supplementary Methods

## 2.1 Fourier smoothing

The normal method of determining strain from two dimensional XRD patterns is a multistep process that firstly bins the data into a set of one-dimensional diffraction patterns, each of equal azimuthal angle ($\psi$). A second step fits the diffraction peaks in each azimuthal slice independently of the other azimuths. The variation in peak centroid ($d_{\psi}^{hkl}$) with azimuth is fit with a $\cos^2\psi$ function that describes the differential strain. This method can be referred to in 2.2.1.

The diffraction patterns in this study present difficulties for this method of XRD fitting. This is due to the small beam spot size (100 $\mu m\times$ 100 $\mu m$), during the DED-AM process the samples results in a grain-size that is a significant fraction of the diffraction volume and the Debye-Scherer rings consist of ∼50-100 very intense diffraction spots overlying a significantly weaker powder background. The spottiness of the IN718 diffraction rings gives scattered peak-centers due to the large variation in maximum intensity and so difference sensitivities to the background and noise in the diffraction pattern, which results in poor constraint in differential strain. The $\gamma$ peaks are also discontinuous but they are also very weak due to the small phase fraction (the brightest peak has an intensity ∼10 on top of a background of ∼40 arbitrary units).

To overcome these problems we utilised a new method for fitting diffraction peaks from 2D data, as a single step process. We assume that the parameters describing a simple Pseudo-Voigt diffraction peak (d-spacing, amplitude, width, peak-shape, etc.) vary smoothly as a function of azimuth and parameterise the variation in each as a Fourier series in azimuth. Thus, as an example, the d-spacing ($d_{\psi}^{hkl}$) is a $2^{nd}$ order Fourier series:

$$d_{\psi}^{hkl} = a_1^{hkl} + b_1^{hkl} + b_2^{hkl}\cos\psi + c_1^{hkl}\sin^2\psi + c_2^{hkl}\cos^2\psi \tag{1}$$

Where $a_1^{hkl}$ is the mean d-spacing, two $1^{st}$ order terms ($b_1$ and $b_2$) represent an offset of the diffraction volume relative to the calibration and are usually very small and the two $2^nd$ order terms ($c_1$ and $c_2$) represent the differential strain. The other peak parameters have more (amplitude) or less (width, peak-shape) terms depending on the requirements of the data.

The model is fit to the 2-dimensional diffraction data at all azimuths within a narrow range around each peak by minimising the least squares residuals. For the $\gamma$ Debye-rings, this function is still leveraged towards the brightest spots in the diffraction rings but the continuity of the d-spacing function ensures that no single diffraction spot dominates the solution. The continuity also ensures that good fits can be gained for the MC carbides data which are otherwise swamped by random noise in the separate azimuthal fits.

## 2.2 Diffraction Analysis Method

### 2.2.1 Determination of macromechanical stress

The following method for determining in-plane stresses is detailed more fully elsewhere [1], which is based on a generalised theory for X-ray determination of stresses from thin films [2]. The details given here apply these methods to the crystallography of the $\gamma$ phase and the sector-by-sector treatment of this dataset. The micromechanical strain can be obtained with knowledge of the lattice plane d-spacing, $d_{\psi}^{hkl}$, for a given $hkl$ reflection and azimuthal angle $\psi$ and a strain free lattice parameter, $d_0^{hkl}$. This is given by

$$\varepsilon_{\psi}^{hkl} = \frac{d_{\psi}^{hkl} - d_0^{hkl}}{d_0^{hkl}} \tag{2}$$

By neglecting small out of plane components, the micromechanical strain can be expressed as

$$\varepsilon_{\psi}^{hkl} = p_{xx,\psi}^{hkl}\sigma_{xx} + p_{xy,\psi}^{hkl}\sigma_{xy} + p_{yy,\psi}^{hkl}\sigma_{yy} \tag{3}$$

where $p_{ij,\psi}^{hkl}$ are lattice plane dependent stress factors and $\sigma_{ij}$ are the 2D components of the macromechnical stress tensor. The stress factors can be expressed in terms of X-ray elastic constants $S_1^{hkl}$ & $\frac{1}{2}S_2^{hkl}$:

$$p_{xx,\psi}^{hkl} = S_1^{hkl} + \frac{1}{2}S_2^{hkl}\cos^2(\psi) \tag{4}$$

$$p_{xy,\psi}^{hkl} = \frac{1}{2}S_2^{hkl}\sin(2\psi) \tag{5}$$

$$p_{yy,\psi}^{hkl} = S_1^{hkl} + \frac{1}{2}S_2^{hkl}\sin^2(\psi) \tag{6}$$

The X-ray elastic constants can be related to the elastic modulus, $E_{hkl}$ and Poisson's ratio, $\nu_{hkl}$ , for a given $hkl$, or in terms of single crystal elastic constant coefficients, $\alpha$, $\beta$, $\eta$ & $\varphi$:

$$S_1^{hkl} = -\frac{\nu^{hkl}}{E^{hkl}} = \eta\Gamma^{hkl} + \varphi \tag{7}$$

$$\frac{1}{2}S_2^{hkl} = \frac{1+\nu^{hkl}}{E^{hkl}} = \Gamma^{hkl}(\alpha - \eta) + \beta - \varphi \tag{8}$$

where $\Gamma^{hkl}$ is the crystallographic factor for a cubic system:

$$\Gamma^{hkl} = \frac{h^2k^2 + h^2l^2 + k^2l^2}{(h^2+k^2+l^2)^2} \tag{9}$$

The high energy diffraction measurements give rise to scattering vectors that lie approximately in a single. The collected 2D Debye-Scherrer diffraction data can therefore be used to calculate the in-plane components of stress. This method inhibits any out of plane stress components, however, neglecting these terms is reasonable as the measured sample was thin. Within a single probed sample location, the $\sigma_{xx}$, $\sigma_{xy}$ & $\sigma_{yy}$ terms can be calculated. These values were obtained from the micromechanical strains (Eq. 2) measured from the first 5 $\gamma_{hkl}$ reflections (111, 200, 220, 311 & 222), at 36 azimuths ($\bar{\psi}_1 = 0°$, $\bar{\psi}_2 = 10°$, $\bar{\psi}_3 = 20°$,..., $\bar{\psi}_{36} = 350°$), totalling 180 measurements. For the calculation of the macromechanical stress tensor, the micromechanical strain, given in Eq. 3, is rewritten as follows:

$$\{\boldsymbol{\varepsilon}\} = \{\boldsymbol{\sigma}\}[\boldsymbol{p}] \tag{10}$$

where

$$\{\boldsymbol{\varepsilon}\} = \{\ \varepsilon^{111}_{\bar{\psi}_1}\ \ \varepsilon^{200}_{\bar{\psi}_1}\ \ \varepsilon^{220}_{\bar{\psi}_1}\ \ \varepsilon^{311}_{\bar{\psi}_1}\ \ \varepsilon^{222}_{\bar{\psi}_1}\ \ \varepsilon^{111}_{\bar{\psi}_2}\ \ \cdots\ \ \varepsilon^{222}_{\bar{\psi}_{36}}\ \}$$

$$\{\boldsymbol{\sigma}\} = \{\ \sigma_{xx}\ \ \sigma_{xy}\ \ \sigma_{yy}\ \}$$

$$[\boldsymbol{p}] = \begin{bmatrix} p^{111}_{xx,\bar{\psi}_1} & p^{200}_{xx,\bar{\psi}_1} & p^{220}_{xx,\bar{\psi}_1} & p^{311}_{xx,\bar{\psi}_1} & p^{222}_{xx,\bar{\psi}_1} & p^{111}_{xx,\bar{\psi}_2} & \cdots & p^{222}_{xx,\bar{\psi}_{36}} \\ p^{111}_{xy,\bar{\psi}_1} & p^{200}_{xy,\bar{\psi}_1} & p^{220}_{xy,\bar{\psi}_1} & p^{311}_{xy,\bar{\psi}_1} & p^{222}_{xy,\bar{\psi}_1} & p^{111}_{xy,\bar{\psi}_2} & \cdots & p^{222}_{xy,\bar{\psi}_{36}} \\ p^{111}_{yy,\bar{\psi}_1} & p^{200}_{yy,\bar{\psi}_1} & p^{220}_{yy,\bar{\psi}_1} & p^{311}_{yy,\bar{\psi}_1} & p^{222}_{yy,\bar{\psi}_1} & p^{111}_{yy,\bar{\psi}_2} & \cdots & p^{222}_{yy,\bar{\psi}_{36}} \end{bmatrix}$$

The values of $p^{hkl}_{ij,\psi}$ given above were calculated from Eq. 4, 5 & 6, where the X-ray elastic constants were obtained using Eq. 7, 8 & 9. Here, experimentally determined single crystal elastic constant coefficients, as a function of temperature, reported for Inconel 718 were used. This single crystal elastic constant coefficients as a function of temperature were fitted with a second order polynomial and extrapolated for the temperature range required for this study. Finally, the macromehanical stresses were determined by rearranging Eq. 10:

$$\{\boldsymbol{\sigma}\} = \{\varepsilon\}[\boldsymbol{p}]^{+} \tag{11}$$

where $[\boldsymbol{p}]^{+}$ is the Moore-Penrose pseudo inverse of $[\boldsymbol{p}]$.
For comparison to yield stress values, the Von Mises stress, $\sigma_{\mathrm{VM}}$ , was calculated from the macromechancial stress tensor. For the in-plane stresses, where $\sigma_{33} = \sigma_{13} = \sigma_{23} = 0$, the Von Mises stress equation is:

$$\sigma_{\mathrm{VM}} = \sqrt{\sigma_{11}^2 + \sigma_{22}^2 + \sigma_{11}\sigma_{22} + 3\sigma_{12}^2} \tag{12}$$

### 2.2.2 Error Calculations

For a given $hkl$ reflection, the d-spacing fitting function outputs 5 coefficients, namely $a_1^{hkl}$, $b_1^{hkl}$, $b_2^{hkl}$, $c_1^{hkl}$, and $c_2^{hkl}$, respectively. The d-spacing at a certain azimuth ($\bar{\psi}_i = 0°, 10°, 20°, ..., 350°$) can be calculated as:

$$d^{hkl}(\bar{\psi}_i) = a_1^{hkl} + b_1^{hkl}\ \cdot\ \sin(\bar{\psi}_i) + b_2^{hkl}\ \cdot\ \cos(\bar{\psi}_i) + c_1^{hkl}\ \cdot\ \sin(2\bar{\psi}_i) + c_2^{hkl}\ \cdot\ \cos(2\bar{\psi}_i) \tag{13}$$

The errors in the lattice d-spacing fitting can propagate through Eq. 13 as:

$$\delta d^{hkl}(\bar{\psi}_i) = \sqrt{(\delta a_1^{hkl})^2 + (\delta b_1^{hkl}\sin(\bar{\psi}_i))^2 + (\delta b_2^{hkl}\cos(\bar{\psi}_i))^2 + (\delta c_1^{hkl}\sin(2\bar{\psi}_i))^2 + (\delta c_2^{hkl}\cos(2\bar{\psi}_i))^2} \tag{14}$$

The errors from the lattice $d$-spacing fitting are propagated in the micromechanical strain calculation through Eq. 14 as:

$$\delta\varepsilon^{hkl}(\bar{\psi}_i) = \sqrt{\left(\frac{\delta d^{hkl}(\bar{\psi}_i)}{d_0^{hkl}}\right)^2 + \left(-\frac{d^{hkl}(\bar{\psi}_i)}{(d_0^{hkl})^2}\ \cdot\ \delta d_0^{hkl}\right)^2} \tag{15}$$

where $d_0^{hkl}$ is a strain free lattice parameter for a given $hkl$ reflection and for a given azimuthal angle $\bar{\psi}_i$.

Expanding Eq. 11 gives

$$\sigma_{ij} = \varepsilon^{111}_{\bar{\psi}_1}\ \cdot\ p^{+}_{1i} + \varepsilon^{200}_{\bar{\psi}_1}\ \cdot\ p^{+}_{2i} + ... + \varepsilon^{222}_{\bar{\psi}_{36}}\ \cdot\ p^{+}_{ni} \tag{16}$$

where $\sigma_{ij} = \sigma_{xx}, \sigma_{xy}$ or $\sigma_{yy}$. The errors in the calculated micromechanical stress terms from Eq. 16 is:

$$\delta\sigma_{ij} = \sqrt{(p_{1i} + \delta\varepsilon^{111}_{\bar{\psi}_1})^2 + (p_{2i} + \delta\varepsilon^{200}_{\bar{\psi}_1})^2 + ... + (p_{ni} + \delta\varepsilon^{222}_{\bar{\psi}_{36}})^2} \tag{17}$$

Finally, the errors in macromechanical stresses calculation from the fitting errors were determined by:

$$\{\delta\boldsymbol{\sigma}\} = \sqrt{\{\delta\boldsymbol{\varepsilon}\}^2[\boldsymbol{p}]^{+^2}} \tag{18}$$

The error in the Von Mises values presented in this work include error contributions from each $\sigma_{ij}$ term, denoted as $\delta\sigma_{ij}$. This is calculated from the following expression

$$\delta\sigma_{\mathrm{VM}} = \sqrt{\left(\frac{2\sigma_{11}+\sigma_{22}}{2\sigma_{\mathrm{VM}}}\cdot\delta\sigma_{11}\right)^2+\left(\frac{2\sigma_{22}+\sigma_{11}}{2\sigma_{\mathrm{VM}}}\cdot\delta\sigma_{22}\right)^2+\left(\frac{3\sigma_{12}}{\sigma_{\mathrm{VM}}}\cdot\delta\sigma_{12}\right)^2} \tag{19}$$

## 2.3 Normal strain and strain rate

In this study, the measured macromechanical stresses were compared to published values of yield strength for the tested material, IN718. However, yield strengths are extremely sensitive to strain rate, making direct comparison without knowledge of deformation rate dependency difficult. Estimations of the strain rate were achieved from the temporally measured stresses. The accumulation of elastic stresses was directly measured here as a function of time, necessitating them to be converted to strains. Specifically, these were calculated for the principal strains, obtained from the following elastic strain relationship from the calculated principal stresses:

$$\begin{pmatrix}\varepsilon_x\\ \varepsilon_y\\ \varepsilon_z\end{pmatrix} = \begin{pmatrix}\frac{1}{E} & -\frac{1}{E} & 0\\ -\frac{v}{E} & \frac{1}{E} & 0\\ 0 & 0 & \frac{1}{G}\end{pmatrix}\begin{pmatrix}\sigma_x\\ \sigma_y\\ 0\end{pmatrix} \tag{20}$$

where $E$ is the Young's modulus, $v$ is the Poisson's ratio, and the shear modulus of elasticity is given by $G = E/2(1+v)$. The principal strains $\varepsilon_x$, $\varepsilon_y$ & $\varepsilon_z$ were calculated from the principal stresses $\sigma_x$ & $\sigma_y$. The latter were obtained from eigenvalues determined from the stress matrix populated from Equation 16. From this calculation, the strain rate was calculated to be ∼0.001 $\mathrm{s}^{-1}$. Yield stress values shown in Fig. 7 of the manuscript were obtained at this strain rate, as a function of temperature from values found elsewhere [3].

## 2.4 Volume fraction determination

The following methodology is adapted from a volume fraction calculation performed elsewhere [4]. The volume fraction of a phase can be deduced from a general expression for the integrated intensity of a diffraction reflection [5]:

$$I^{\alpha}_{hkl} = \left[\frac{I_0\lambda^3}{32\pi r}\frac{e_c^4}{m_e c^4}\right]\left[\frac{p_{hkl}}{2V_\alpha^2}\cdot|F^2_{hkl}|\cdot\frac{1+\cos^2 2\theta\cos^2 2\theta_m}{\sin^2\theta\cos\theta}\right]\left[\frac{W_\alpha}{\rho_\alpha\mu^*_m}\right] \tag{21}$$

where $I_0$ is the incident X-ray beam intensity, $\lambda$ is the wavelength, $r$ is the distance between the detector and a scattering electron, $e_c$ is the electron charge, $m_e$ is the mass of an electron, $c$ is the speed of light, $p_{hkl}$ is the multiplicity of a reflection $hkl$, $V_\alpha$ is the volume of the unit cell, $F_{hkl}$ is the structure factor, $\theta$ is the diffraction angle for reflection $hkl$ and $\theta_m$ is the monochromator angle. $W_\alpha$ and $\rho_\alpha$ are the weight fraction and density of phase $\alpha$, and $\mu^*_m$ is the sample mass absorption coefficient. Many of these terms do not need to be known in order to ascertain a phase volume fraction, as they are simply scaling constants. Instead it can be deduced from the relative integrated intensity , denoted here as $I^{\mathrm{R}}_{(hkl)}$:

$$I^{\mathrm{R}}_{hkl} = I_0 p_{hkl}|F^2_{hkl}|\frac{1+\cos^2 2\theta}{2}\cdot\frac{1}{\sin\theta\sin 2\theta}A(\theta)e^{-2M_T} \tag{22}$$

where $I_0$ is the incident X-ray beam intensity, $(1+\cos^2 2\theta)/2$ is a polarisation factor, equal to 1 for synchrotron radiation, $1/(\sin\theta\sin 2\theta)$ is the Lorenz factor, $A(\theta)$ is the diffraction angle dependent absorption constant, and $e^{(-2M_T)}$ is the Debye-Waller temperature factor.

The term $W_\alpha/\rho_\alpha$ from Eq 21 is equal to the volume fraction of this phase, $\phi_\alpha$. The polarisation factor is equal to 1 in synchrotron radiation. As the reflections assessed over a narrow $2\theta$ range, the angle dependency of the absorption constant is small, so this term can be neglected. Combining these simplifications, Eq 21 and 22, for a hypothetical phase $\alpha$ can be described as:

$$I^{\alpha} = I^{\alpha}_s\cdot p_{hkl}\cdot|F^2_{hkl}|\cdot\frac{1}{\sin\theta\sin 2\theta}\cdot e^{(-2M_T)}\phi_\alpha \tag{23}$$

Table 1: Crystallographic details used for structure factor calculations.

| Phase | Space Group | Sublattice 1 Coordinates | Sublattice 2 Coordinates | Notes |
|---|---|---|---|---|
| $\gamma$ | Fm$\bar{3}$m | [0 0 0], [0 $\frac{1}{2}$ $\frac{1}{2}$], [0 $\frac{1}{2}$ $\frac{1}{2}$], [$\frac{1}{2}$, 0 ,$\frac{1}{2}$] | - | A1 crystal structure. |
| MC | Pm$\bar{3}$m | [0 0 0], [0 $\frac{1}{2}$ $\frac{1}{2}$], [0 $\frac{1}{2}$ $\frac{1}{2}$], [$\frac{1}{2}$, 0, $\frac{1}{2}$] | [0 0 $\frac{1}{2}$], [0 $\frac{1}{2}$ 0], [$\frac{1}{2}$ 0 0], [$\frac{1}{2}$ $\frac{1}{2}$ $\frac{1}{2}$] | $L1_2$ crystal structure. Based on TiC compound, where Ti is sublattice 1 and C is sublattice 2. |
| Laves | $P6_3$/mmc | [0 0 0], [0 0 $\frac{1}{2}$], [0.83 0.17 $\frac{1}{4}$], [0.34 0.17 $\frac{1}{4}$], [0.17 0.34 $\frac{3}{4}$], [0.66 0.83 $\frac{3}{4}$], [0.17 0.83 $\frac{3}{4}$], [0.83 0.66 $\frac{1}{4}$] | [$\frac{1}{3}$ $\frac{2}{3}$ $\frac{1}{4}$] [$\frac{1}{3}$ $\frac{2}{3}$ 0.437], [$\frac{2}{3}$ $\frac{1}{3}$ 0.563], [$\frac{2}{3}$ $\frac{1}{3}$ 0.937] | C14 crystal structure. Based on $Fe_2Nb$ compound, where Fe is sublattice 1, and Nb is sublattice 2. |

For a reflection, $hkl$ in a phase, the intensity scales by a term $I_{hkl}$, where

$$I_{hkl} = p_{hkl} \cdot |F_{hkl}^2| \cdot \frac{1}{\sin\theta \sin 2\theta} \cdot e^{(-2M_T)} \tag{24}$$

Thus, it is possible to obtain the volume fraction contributions from each phase by combining Eq 23 and 24:

$$\phi_\gamma = \left(\frac{I^\gamma}{I_{hkl}}\right)\frac{1}{I_s} \tag{25}$$

$$\phi_{\text{Laves}} = \left(\frac{I^{\text{Laves}}}{I_{hkl}}\right)\frac{1}{I_s} \tag{26}$$

$$\phi_{\text{MC}} = \left(\frac{I^{\text{MC}}}{I_{hkl}}\right)\frac{1}{I_s} \tag{27}$$

The only remaining unknown in the above expression is no $I_s$. This can be simply calculated by summing each volume fraction expression:

$$\frac{I^\gamma}{I_{hkl}} + \frac{I^{\text{Laves}}}{I_{hkl}} + \frac{I^{\text{MC}}}{I_{hkl}} = I_s(\phi_\gamma + \phi_{\text{Laves}} + \phi_{\text{MC}}) \tag{28}$$

as $\phi_\gamma + \phi_{\text{Laves}} + \phi_{\text{MC}} = 1$, $I_s$ is known and hence the individual volume fractions can be calculated.

### 2.4.1 Structure Factor

For each $hkl$ reflection, and for each phase, a structure factor must be calculated. The structure factors here are strongly related to their phase compositions, and more specifically, their site occupancies. Expressions for the structure factors of each phase was determined to account for this. The disordered A1 structure $\gamma$ phase has a single sublattice and is given by the following general solution:

$$F_{hkl}^\gamma = \sum c_j f_j e^{[-2\pi i(hx_j + ky_j + lz_j)]} \tag{29}$$

where $c_j$ is the site occupancy for each element $j$, on the corresponding lattice coordinates $x_j$, $y_j$ and $z_j$, and $f_j$ is the atomic scattering factor, for element $j$. For the MC carbides and Laves phases, they have two sublattices, denoted here as $a$ & $b$. As each element, $j$ may not reside on a single sublattice, the structure factors for each site can be treated independently as follows:

$$F_{hkl} = \sum_a c_{j_a} f_j e^{[-2\pi i(hx_{j_a} + ky_{j_a} + lz_{j_a})]} + \sum_b c_{j_b} f_j e^{[-2\pi i(hx_{j_b} + ky_{j_b} + lz_{j_b})]} \tag{30}$$

where $c_{j_a}$ & $c_{j_b}$ are the site occupancies of element $j$ on sublattice $a$ and $b$. The lattice coordinates $x_{j_a}$, $y_{j_a}$ and $z_{j_a}$ are assigned per element $j$ on sublattice $a$ and lattice coordinates $x_{j_b}$, $y_{j_b}$ and $z_{j_b}$ are assigned per element $j$ on sublattice $b$. The sublattice coordinates for the phases calculated are given in Table 1.

The general expression for the atomic scattering factors, $f$, of any element is defined as

$$f = f_0 + \Delta f' + i\Delta f'' \tag{31}$$

where $f_0$ is the real atomic scatting factor, and the complex $\Delta f' + i\Delta f''$ term is a dispersion correction. The real atomic scattering factor, valid for the range $0 < \sin\theta/\lambda < 2.0\,\text{Å}$ is as follows:

$$f_0 = \sum_{i=1}^{4} a_i e^{(-b_i \sin^2\theta/\lambda^2)} + c \tag{32}$$

where $a_i$, $b_i$ and $c$ are element dependent coefficients that are tabulated elsewhere [6]. The dispersion correction terms can be found in Ref [7].

### 2.4.2 Thermodynamic modelling

The structure factor calculation, Eq. 29 & 30, requires knowledge of the crystal structure site occupancies, as a function of temperature. These were predicted with Thermo-Calc [8] v2020a, using the TCNi8 (v8.2) database. In addition to the presence of liquid, from identification of the reflections from the diffraction patterns, the solid state phases were known to be present were $\gamma$, MC carbides and Laves. The simulations included these phases, and suspended all others. For crystalline phases that formed whilst solidification was talking place, a Scheil solidification simulation with back diffusion was used. This simulation incorporates the secondary dendrite arm spacing (SDAS), which is related empirically to the cooling rate, $\mathrm{d}T/\mathrm{d}t$ (via $\mathrm{SDAS} = A(\mathrm{d}T/\mathrm{d}t)^{-n}$, where $A$ and $n$ are constants. This function was fitted to published data on Alloy 718 [9], then the ThermoCalc simulation was run with $\mathrm{d}T/\mathrm{d}t = 500\,\mathrm{Ks^{-1}}$. For predictions below the solidus temperature, the site occupancy predictions were based on equilibrium simulations.

### 2.4.3 Temperature factor

The variation in intensity between each reflection for a given phase is governed by each term in Eq 24. The multiplicity, structure factor and Lorenz factor can be calculated, however, the Debye temperature factor $e^{-2M_T}$ remains unknown, and must be calculated from the data. It is known that

$$2M_T = \frac{16\pi^2\langle u_s^2\rangle(\sin^2\theta)}{\lambda^2} \tag{33}$$

where $u_s$ is the atomic displacement normal to the diffracting plane. For a given phase, if one takes the integrated intensity of several $hkl$ reflection and corrects them with the multiplicity, structure factor and Lorenz factor (this will be called $I^*_{hkl}$), any difference between the different $hkl$ intensities should differ only by their Deybe temperature factor. Using this property, the log of the $I^*_{hkl}$ reflection intensities was plotted against $\sin^2\theta/\lambda$. Fitting a linear equation to this data yields a gradient of $16\pi^2\langle u_s^2\rangle$, enabling the Debye temperature correction to be made. This method assumes $\langle u_s^2\rangle$ does not vary with $hkl$.

## 2.5 Solidification microstructure calculation

The microstructure of the solidification front can be determined using the following equations [10]:

$$V_c = \frac{GD}{\Delta T_0} \tag{34}$$

$$V_{c-d} = \frac{GD}{k\Delta T_0} \tag{35}$$

Where $V_c$ is the lower velocity threshold for planar front solidification and $V_{c-d}$ is the lower velocity threshold for the cellular-dendritic transition. $G$ is the mean temperature gradient at the interface.$D$ is the diffusion coefficient. $\Delta T_0$ represents the equilibrium liquidus/solidus difference. k is the partition coefficient.

### 2.6 Marangoni convection calculation

The Marangoni convection can be determined by a dimensionless number[11]:

$$M_a = \frac{d\gamma_s}{dT}\frac{dT}{dx}\frac{L^2}{\eta\alpha} \tag{36}$$

Where $M_a$ is the dimensionless Marangoni number, $\gamma_s$ is the surface tension, $\frac{dT}{dx}$ is the temperature gradient, $\alpha$ is the thermal diffusivity, $L$ is the characteristic length and $\eta$ is the viscosity of the melt pool .

### 2.7 Weber number and Péclet number calculation

To describe the kinetic energy in the melt pool, the Weber number ($W_e$) is adopted in our study. The Weber number is used to describe the flow damping on the surface [12],

$$W_e = \frac{\rho v^2 L}{\sigma} \tag{37}$$

where $\rho$ is the liquid density, $v$ is the flow speed, $L$ is the characteristic length, and $\sigma$ is the surface tension. To describe heat transfer in the melt pool, the Péclet number ($Pe$) which combines the Reynolds number ($Re_L$) and Prandtl number ($Pr$)[12], was used as,

$$Pe = Re_L Pr = \frac{\rho v L}{\mu}\frac{c_p\mu}{k} = \frac{Lv}{\alpha} \tag{38}$$

where $L$ is the characteristic length, $v$ is the local flow speed, and $\alpha$ is the thermal diffusivity.

### 2.8 Melt pool volume calculation

To calculate the melt pool volume, an assumption is made that the melt pool of Inconel718 during DED-AM is a tear-drop shape. Therefore, the melt-pool can be divided as two semi-elliptical caps. The schematic of the melt pool geometry measurements is shown below:

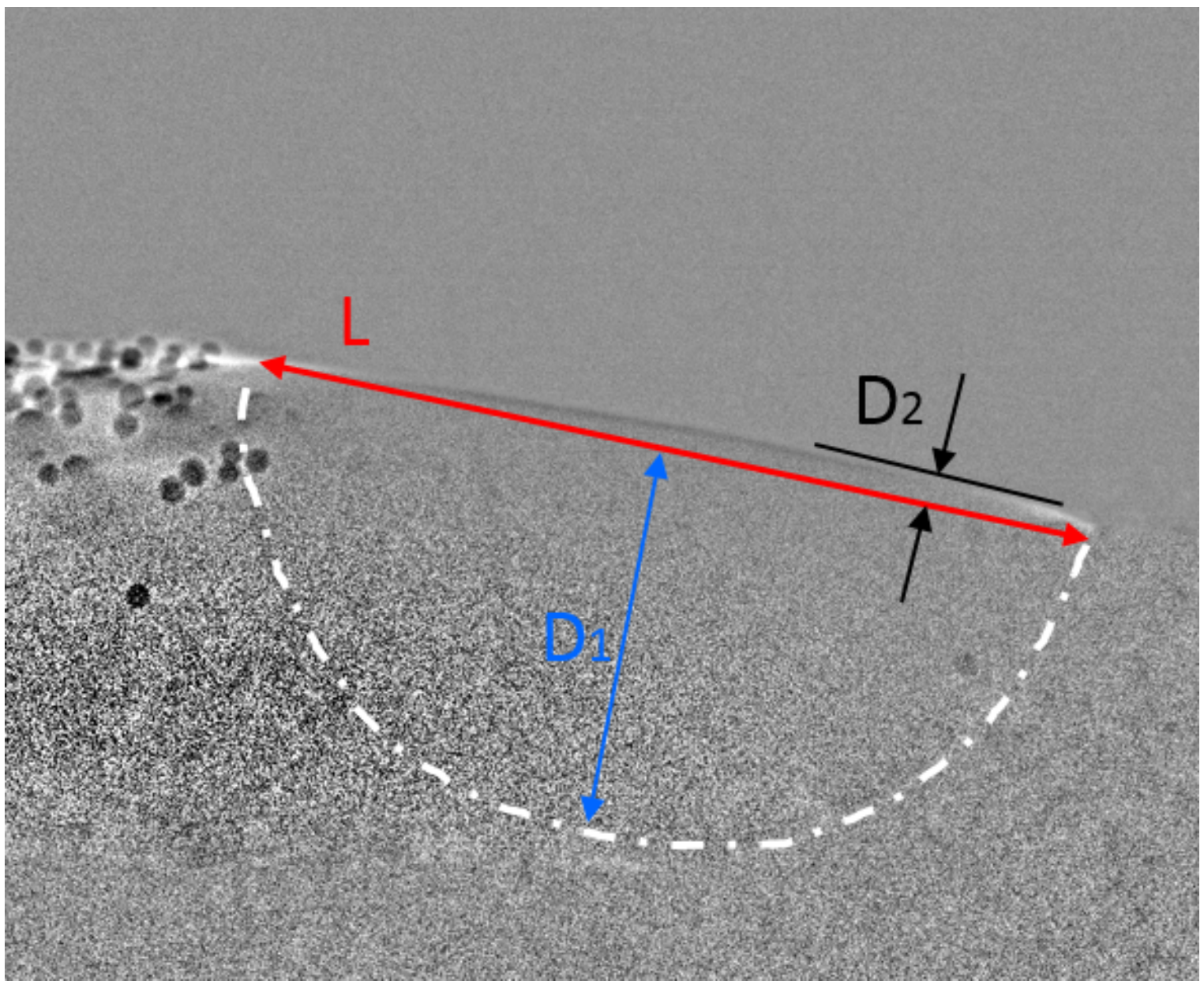

Figure 1: **Melt pool geometry measurements.** $L$ is the melt pool length, $D_1$ and $D_2$ are the melt pool depth. The width of the melt pool ($W$) is set to be 1 $mm$, which is the thickness of the substrate.

The melt pool volume measurements follow the below equation:

$$V_{meltpool} = \frac{\pi(L/2)W}{3(L/2)^2}D_2^2(3(L/2) - D_2) + \frac{\pi(L/2)W}{3(L/2)^2}D_1^2(3(L/2) - D_1) \tag{39}$$

# 3 Supplementary Results

## 3.1 Marangoni convection

### 3.1.1 *In situ* tracer velocity measurements

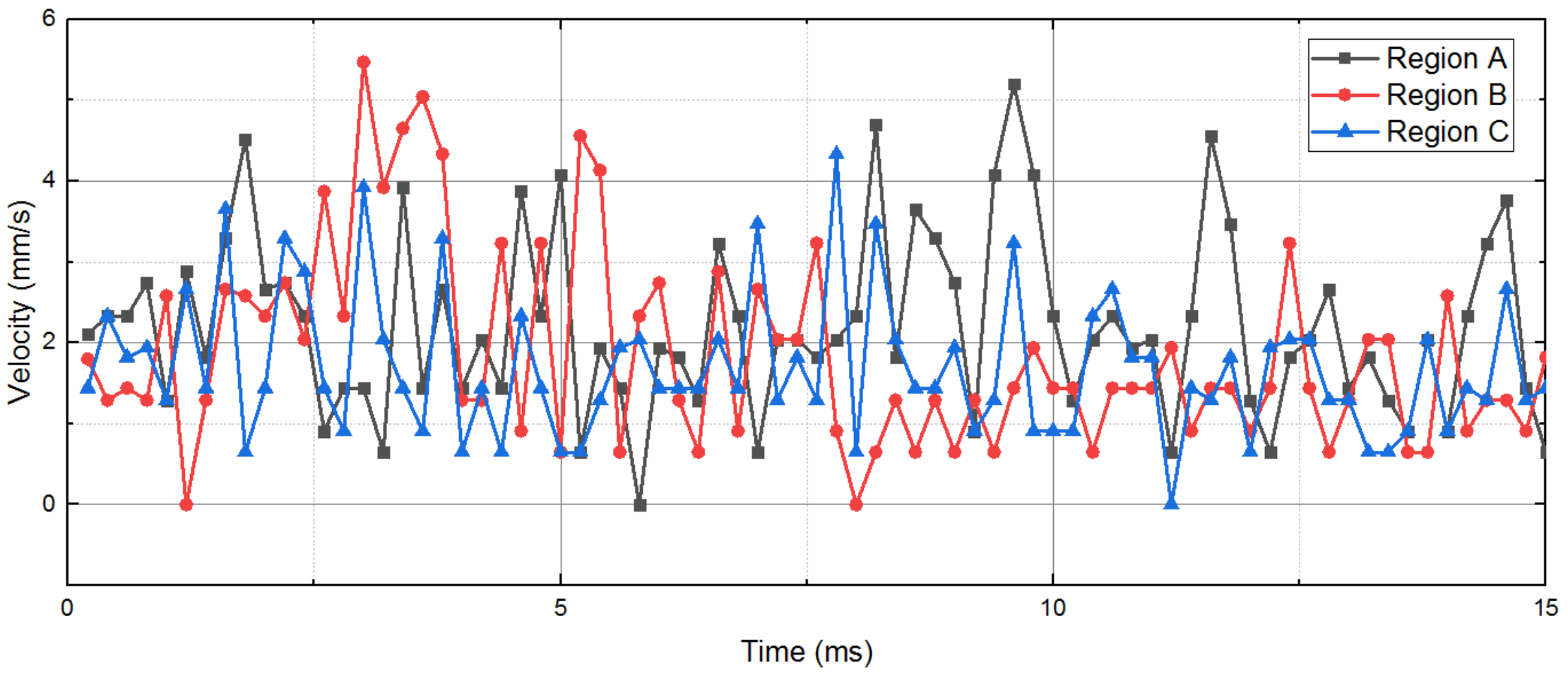

Figure 2: **Instantaneous tracer velocity measurements.** The instant speed of W tracers in region A, B and C were carefully measured by calculating the displacement of W tracers over each frame of the radiography. The max speed of the W tracers in the melt pool is 5.5 $\mathrm{mm\,s^{-1}}$ and the average speed in all regions is 2 $\mathrm{mm\,s^{-1}}$.

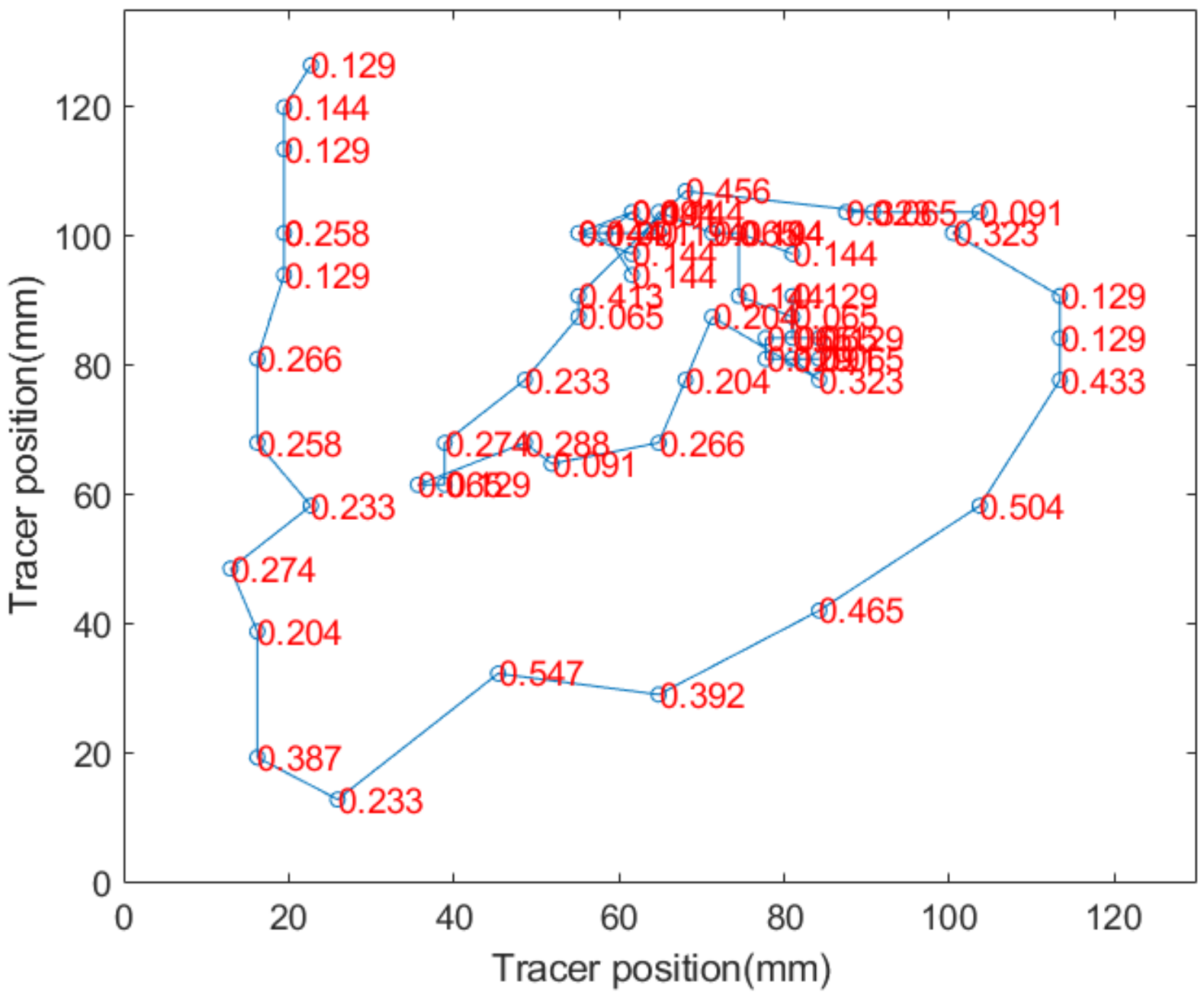

Figure 3: **Instantaneous tracer velocity measurements against its instantaneous position.** The instant speed of W tracers in region C were carefully measured by calculating the displacement of W tracers over each frame of the radiography. It is plotted against its instantaneous position on the radiograph, the unit is mm. The unit for the tracer velocity is $\mathrm{cm\,s^{-1}}$.

### 3.1.2 The projected average speed measurements

| Melt pool region | 1 mm s$^{-1}$ Layer 1 | 1 mm s$^{-1}$ Layer 2 | 1.5 mm s$^{-1}$ Layer 1 | 1.5 mm s$^{-1}$ Layer 2 |
|---|---|---|---|---|
| **A** | 2.05 | 1.94 | 1.86 | 1.92 |
| **B** | 2.34 | 2.21 | 2.57 | 1.61 |
| **C** | 1.72 | 2.23 | 2.04 | 1.77 |

Table 2: **The projected average speed of W tracers in the melt pool.** The instantaneous speed of W tracers layer 1 and layer 2 of build speed of 1 mm s$^{-1}$ and 2 mm s$^{-1}$ are measured respectively and the average speed was calculated.

### 3.1.3 Weber number ($W_e$) and Péclet number ($P_e$)

The characteristic length of the melt pool centre ($L_c$) is chosen as the melt pool depth, which is 2 mm, while the characteristic length of the solidification front ($L_s$) is chosen as the melt pool depth at the melt pool boundary, which is estimated to be ~50 µm. $\rho_f$ is calculated to be 7529 kg m$^{-3}$ and $\rho_c$ is estimated to be 7134 kg m$^{-3}$ [13]. $\nu_f$ is measured to be 1 mm s$^{-1}$ while $\nu_c$ is 3 mm s$^{-1}$. The surface tension at liquidus temperature is ~1970 mN m$^{-1}$, while in the melt pool centre is 1487 nN m$^{-1}$ [14]. The liquidus temperature at the solidification front is set to be 1360 K in our calculation and the temperature at the melt pool center is set to be 2000 $K$. The thermal diffusivity, $\alpha_s$ is 0.055903 cm$^2$ s$^{-1}$ while $\alpha_f$ is 0.0387638 cm$^2$ s$^{-1}$ [15] Based on these parameters, the Weber number is calculated to be 43 near the melt pool center and 0.19 near the melt pool tail. And the $P_e$ number is calculated to be 107.3 near the melt pool center and 1.29 at the solidification front.

## 3.2 Diffraction peaks index

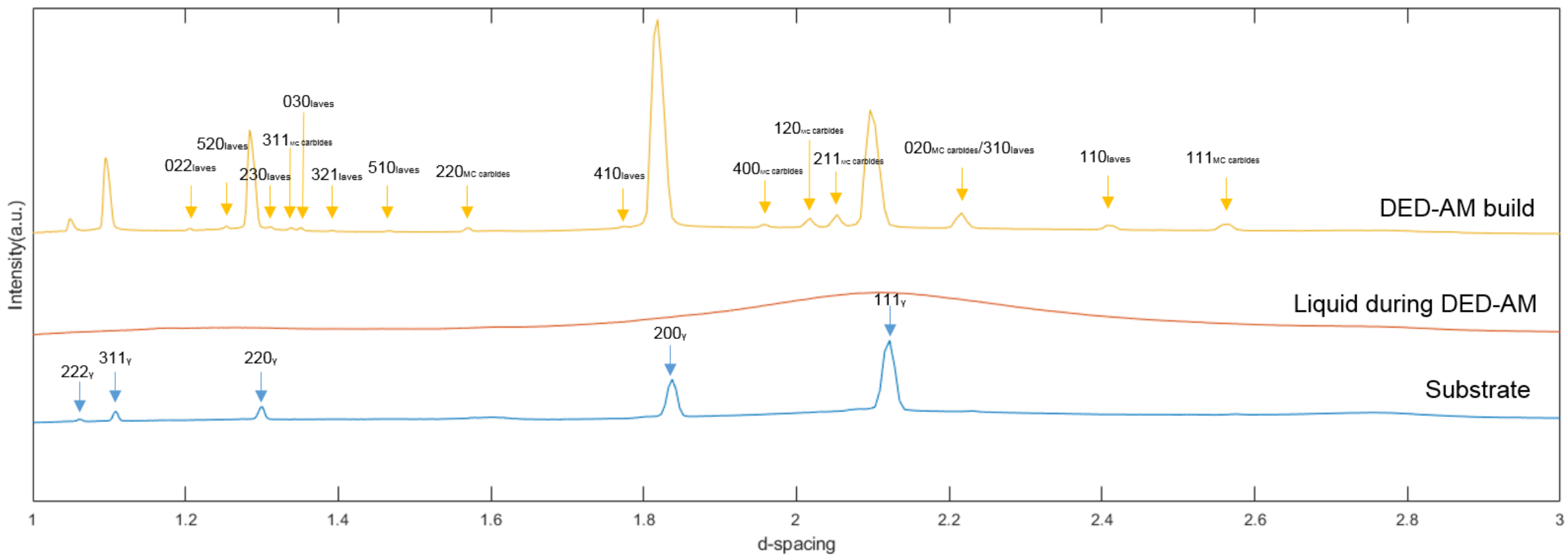

Figure 4: **Peak index.** $\gamma$, MC carbides and Laves phases are identified in the diffraction pattern.

## 3.3 Melt pool mapping

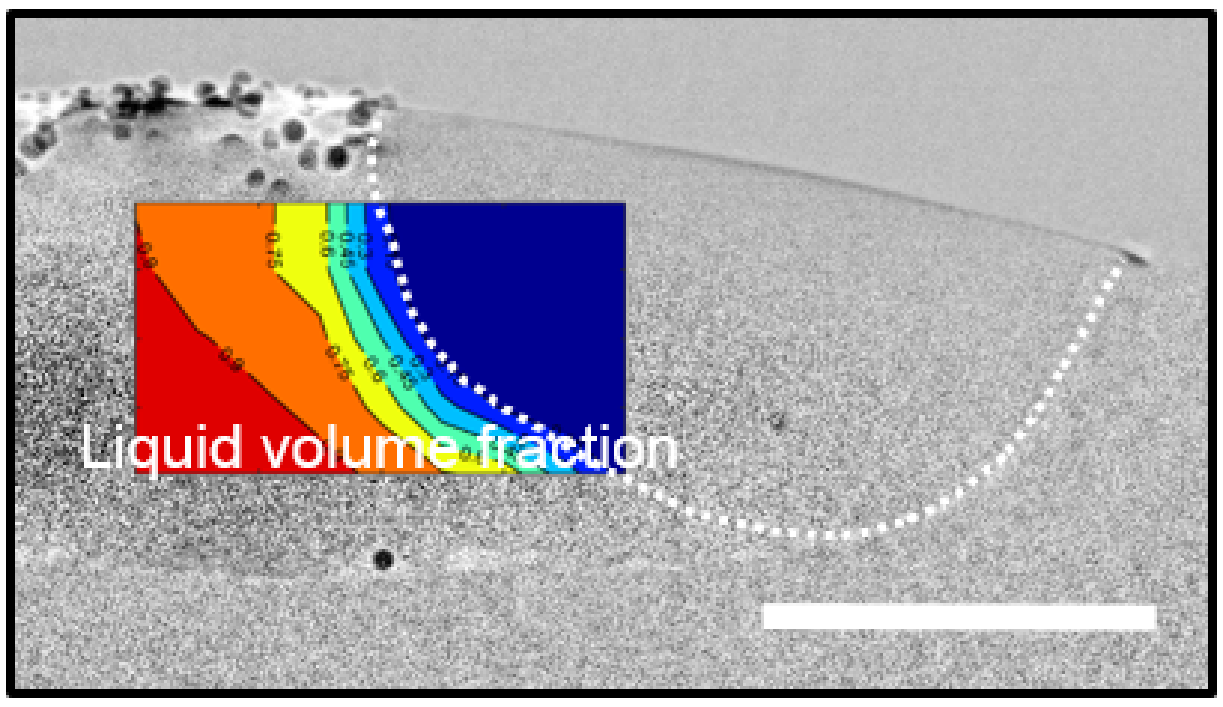

Figure 5: **Liquid volume fraction calculated from diffraction patterns matches the melt pool boundary captured using X-ray radiography.** Liquid volume fraction results are interpolated and show perfect gradient matching the radiography results. Scale bar = 1000 $\mu m$.

| | | | | | | | | |
|---|---|---|---|---|---|---|---|---|
| **Temperature calculation - Layer 1** (°C) | | | | | | | | |
| 1254.541 | 1278.755 | 1278.364 | 1278.331 | 1329.431 | 1360 | 1360 | 1360 | 1360 |
| 1246.216 | 1272.723 | 1287.459 | 1279.226 | 1316.956 | 1360 | 1360 | 1360 | 1360 |
| 1244.378 | 1266.281 | 1263.335 | 1281.096 | 1276.102 | 1335.996 | 1360 | 1360 | 1360 |
| 1229.982 | 1259.989 | 1265.066 | 1278.470 | 1278.572 | 1288.147 | 1320.453 | 1360 | 1360 |
| 1198.320 | 1237.145 | 1254.965 | 1261.543 | 1292.129 | 1305.663 | 1362.137 | 1366.400 | 1334.694 |
| **Temperature calculation - Layer 3** (°C) | | | | | | | | |
| 1256.488 | 1276.861 | 1282.929 | 1286.663 | 1286.663 | 1286.663 | 1360 | 1360 | 1360 |
| 1251.096 | 1271.869 | 1284.200 | 1282.503 | 1282.503 | 1282.503 | 1360 | 1360 | 1360 |
| 1243.212 | 1272.836 | 1269.927 | 1282.888 | 1282.888 | 1282.888 | 1360 | 1360 | 1360 |
| 1223.447 | 1258.195 | 1255.387 | 1270.919 | 1270.919 | 1270.919 | 1263.638 | 1360 | 1360 |
| 1193.526 | 1251.187 | 1252.027 | 1261.385 | 1261.385 | 1261.385 | 1332.913 | 1320.980 | 1249.759 |
| **Temperature calculation - Layer 5** (°C) | | | | | | | | |
| 1272.381 | 1277.358 | 1281.218 | 1285.488 | 1301.328 | 1360 | 1360 | 1360 | 1360 |
| 1247.802 | 1256.164 | 1271.380 | 1286.518 | 1302.936 | 1360 | 1360 | 1360 | 1360 |
| 1248.773 | 1274.634 | 1269.605 | 1276.204 | 1260.301 | 1293.443 | 1360 | 1360 | 1360 |
| 1230.393 | 1259.411 | 1255.603 | 1277.815 | 1289.027 | 1297.570 | 1298.807 | 1309.067 | 1360 |
| 1209.713 | 1255.295 | 1238.127 | 1249.686 | 1275.763 | 1301.372 | 1302.547 | 1343.106 | 1360 |

Table 3: **Temperature data in the melt pool area in layer 1, 3 & 5.** The temperature values are calculated from the lattice spacing measured from the $\gamma$ (200) reflections. IN718 build is under $P$ = 1.59 $\times$ $10^3$ W mm$^{-3}$, $v$ = 1 mm s$^{-1}$, powder feedrate 1 g min$^{-1}$. All diffraction experiments used the same parameters. Temperature in the melt pool is set to be 1360 °C.

| Liquid volume fraction - Layer 1 | | | | | | | | |
|---|---|---|---|---|---|---|---|---|
| 0.040 | 0.144 | 0.212 | 0.336 | 0.830 | 0.843 | 0.606 | 1 | 1 |
| 0.062 | 0.173 | 0.220 | 0.332 | 0.815 | 0.998 | 0.996 | 0.894 | 1 |
| 0.000 | 0.087 | 0.134 | 0.205 | 0.546 | 0.976 | 0.988 | 0.993 | 0.969 |
| 0.000 | 0.045 | 0.073 | 0.122 | 0.292 | 0.640 | 0.874 | 0.994 | 1 |
| 0.000 | 0.011 | 0.047 | 0.061 | 0.115 | 0.233 | 0.293 | 0.632 | 0.905 |
| **Liquid volume fraction - Layer 3** | | | | | | | | |
| 0.000 | 0.000 | 0.192 | 0.279 | 0.839 | 0.898 | 0.599 | 1 | 1 |
| 0.029 | 0.112 | 0.150 | 0.248 | 0.671 | 0.960 | 1 | 0.953 | 1 |
| 0.000 | 0.091 | 0.076 | 0.122 | 0.334 | 0.834 | 0.956 | 0.975 | 0.926 |
| 0.000 | 0.022 | 0.039 | 0.041 | 0.158 | 0.371 | 0.738 | 0.943 | 0.993 |
| 0.000 | 0.000 | 0.000 | 0.005 | 0.024 | 0.106 | 0.159 | 0.446 | 0.953 |
| **Liquid volume fraction - Layer 5** | | | | | | | | |
| 0.000 | 0.000 | 0.116 | 0.300 | 0.672 | 0.793 | 0.751 | 1 | 1 |
| 0.008 | 0.000 | 0.130 | 0.272 | 0.317 | 0.950 | 0.964 | 0.979 | 1 |
| 0.024 | 0.000 | 0.098 | 0.121 | 0.159 | 0.652 | 0.868 | 0.981 | 1 |
| 0.000 | 0.069 | 0.045 | 0.052 | 0.034 | 0.274 | 0.367 | 0.879 | 0.868 |
| 0.000 | 0.022 | 0.000 | 0.021 | 0.050 | 0.088 | 0.038 | 0.277 | 1 |

Table 4: **Molten liquid volume fraction in the melt pool area in layer 1, 3 & 5.** The molten liquid volume fraction are calculated from the liquid reflections.

| $\gamma$ volume fraction - Layer 1 | | | | | | | | |
|---|---|---|---|---|---|---|---|---|
| 0.891 | 0.853 | 0.784 | 0.664 | 0.170 | 0 | 0 | 0 | 0 |
| 0.928 | 0.825 | 0.777 | 0.668 | 0.185 | 0 | 0 | 0 | 0 |
| 0.991 | 0.910 | 0.861 | 0.791 | 0.454 | 0.024 | 0 | 0 | 0 |
| 0.990 | 0.950 | 0.921 | 0.873 | 0.708 | 0.360 | 0.126 | 0 | 0 |
| 0.988 | 0.984 | 0.947 | 0.934 | 0.881 | 0.767 | 0.707 | 0.368 | 0.095 |
| **$\gamma$ volume fraction - Layer 3** | | | | | | | | |
| 0.990 | 0.897 | 0.805 | 0.703 | 0 | 0 | 0 | 0 | 0 |
| 0.967 | 0.882 | 0.846 | 0.742 | 0.324 | 0 | 0 | 0 | 0 |
| 0.990 | 0.904 | 0.919 | 0.874 | 0.663 | 0.165 | 0 | 0 | 0 |
| 0.995 | 0.974 | 0.957 | 0.953 | 0.837 | 0.625 | 0.262 | 0 | 0 |
| 0.992 | 0.990 | 0.993 | 0.989 | 0.971 | 0.889 | 0.832 | 0.554 | 0.047 |
| **$\gamma$ volume fraction - Layer 5** | | | | | | | | |
| 0.943 | 0.903 | 0.882 | 0.605 | 0.328 | 0 | 0 | 0 | 0 |
| 0.982 | 0.923 | 0.864 | 0.725 | 0.683 | 0 | 0 | 0 | 0 |
| 0.972 | 0.996 | 0.899 | 0.874 | 0.841 | 0.348 | 0 | 0 | 0 |
| 0.991 | 0.968 | 0.946 | 0.942 | 0.962 | 0.726 | 0 | 0 | 0 |
| 0.994 | 0.974 | 0.987 | 0.966 | 0.946 | 0.911 | 0.817 | 0.723 | 0 |

Table 5: $\gamma$ **volume fraction in the melt pool area in layer 1, 3 & 5.** The $\gamma$ phase volume fraction are calculated from the $\gamma$ reflections.

| MC carbides volume fraction - Layer 1 | | | | | | | | |
|---|---|---|---|---|---|---|---|---|
| 6.27E-03 | 3.26E-03 | 4.69E-03 | 0 | 0 | 0 | 0 | 0 | 0 |
| 9.28E-03 | 2.34E-03 | 3.01E-03 | 0 | 0 | 0 | 0 | 0 | 0 |
| 8.86E-03 | 3.64E-03 | 4.64E-03 | 3.63E-03 | 0 | 0 | 0 | 0 | 0 |
| 9.76E-03 | 4.54E-03 | 6.78E-03 | 4.71E-03 | 0 | 0 | 0 | 0 | 0 |
| 1.17E-02 | 5.26E-03 | 5.84E-03 | 4.62E-03 | 4.14E-03 | 0 | 0 | 0 | 0 |
| **MC carbides volume fraction - Layer 3** | | | | | | | | |
| 9.58E-03 | 6.38E-03 | 3.19E-03 | 1.74E-02 | 0 | 0 | 0 | 0 | 0 |
| 3.76E-03 | 5.83E-03 | 3.64E-03 | 9.63E-03 | 4.21E-03 | 0 | 0 | 0 | 0 |
| 9.72E-03 | 4.92E-03 | 4.28E-03 | 3.82E-03 | 2.90E-03 | 7.22E-04 | 0 | 0 | 0 |
| 4.86E-03 | 4.34E-03 | 4.60E-03 | 6.03E-03 | 5.30E-03 | 3.95E-03 | 0 | 0 | 0 |
| 7.77E-03 | 9.49E-03 | 6.88E-03 | 5.48E-03 | 5.38E-03 | 4.93E-03 | 9.51E-03 | 0 | 0 |
| **MC carbides volume fraction - Layer 5** | | | | | | | | |
| 6.46E-03 | 4.66E-03 | 2.13E-03 | 0 | 0 | 0 | 0 | 0 | 0 |
| 8.26E-03 | 7.19E-03 | 6.11E-03 | 2.34E-03 | 0 | 0 | 0 | 0 | 0 |
| 3.77E-03 | 4.45E-03 | 2.97E-03 | 5.17E-03 | 0 | 0 | 0 | 0 | 0 |
| 8.36E-03 | 8.76E-03 | 9.16E-03 | 5.43E-03 | 3.26E-03 | 0 | 0 | 0 | 0 |
| 5.53E-03 | 3.86E-03 | 1.29E-02 | 8.39E-03 | 3.84E-03 | 1.39E-03 | 0 | 0 | 0 |

Table 6: **MC carbides volume fraction in the melt pool area in layer 1, 3 & 5.** The MC carbides phase volume fraction are calculated from the MC carbides reflections.

| Laves volume fraction - Layer 1 | | | | | | | | |
|---|---|---|---|---|---|---|---|---|
| 0 | 0 | 0 | 0 | 0 | 0 | 0 | 0 | 0 |
| 3.27E-04 | 0 | 0 | 0 | 0 | 0 | 0 | 0 | 0 |
| 2.32E-04 | 0 | 0 | 0 | 0 | 0 | 0 | 0 | 0 |
| 2.58E-04 | 0 | 0 | 0 | 0 | 0 | 0 | 0 | 0 |
| 1.26E-04 | 2.60E-05 | 0 | 0 | 0 | 0 | 0 | 0 | 0 |
| **Laves volume fraction - Layer 3** | | | | | | | | |
| 1.86E-04 | 0 | 0 | 0 | 0 | 0 | 0 | 0 | 0 |
| 1.07E-04 | 0 | 0 | 0 | 0 | 0 | 0 | 0 | 0 |
| 2.49E-04 | 0 | 0 | 0 | 0 | 0 | 0 | 0 | 0 |
| 1.60E-04 | 0 | 0 | 0 | 0 | 0 | 0 | 0 | 0 |
| 2.78E-04 | 3.48E-04 | 1.51E-04 | 0 | 0 | 0 | 0 | 0 | 0 |
| **Laves volume fraction - Layer 5** | | | | | | | | |
| 1.67E-03 | 0 | 0 | 0 | 0 | 0 | 0 | 0 | 0 |
| 9.80E-04 | 0 | 0 | 0 | 0 | 0 | 0 | 0 | 0 |
| 2.94E-04 | 0 | 0 | 0 | 0 | 0 | 0 | 0 | 0 |
| 2.85E-04 | 0 | 0 | 0 | 0 | 0 | 0 | 0 | 0 |
| 2.88E-04 | 4.24E-04 | 5.60E-04 | 0 | 0 | 0 | 0 | 0 | 0 |

Table 7: **Laves volume fraction in the melt pool area in layer 1, 3 & 5.** The Laves phase volume fraction are calculated from the Laves reflections.

### 3.4 Melt track mapping

| Temperature calculation - Melt track area ($^{\circ}C$) | | | | | | | | |
|---|---|---|---|---|---|---|---|---|
| Line 1 | 1360 | 1360 | 1360 | 1360 | 1360 | 1360 | 1322.954 | 1279.089 |
| Line 2 | 1360 | 1360 | 1360 | 1360 | 1360 | 1293.389 | 1286.023 | 1286.146 |
| Line 3 | 1327.125 | 1314.012 | 1314.451 | 1295.842 | 1289.807 | 1270.561 | 1257.349 | 1241.551 |
| **– – continued** | | | | | | | | |
| Line 1 | 1280.326 | 1252.752 | 1216.351 | 1188.476 | 1168.203 | 1117.131 | 1082.769 | 1058.208 |
| Line 2 | 1268.104 | 1241.212 | 1202.869 | 1141.174 | 1142.112 | 1068.436 | 1036.341 | 1020.542 |
| Line 3 | 1197.231 | 1148.956 | 1133.266 | 1109.442 | 1085.617 | 1038.432 | 1011.193 | 998.935 |
| **– – continued** | | | | | | | | |
| Line 1 | 1016.091 | 992.243 | 1019.205 | 941.779 | 932.929 | 898.288 | 878.966 | 869.111 |
| Line 2 | 1010.625 | 981.511 | 976.681 | 926.045 | 917.949 | 894.174 | 882.446 | 875.211 |
| Line 3 | 974.405 | 923.080 | 951.157 | 903.821 | 892.922 | 873.978 | 870.490 | 826.217 |

Table 8: **Temperature data in the melt track area in the melt track region, re-melt region and reheat region.** The temperature values are calculated from the lattice spacing measured from the $\gamma$ (200) reflections.

| Liquid volume fraction calculation - Melt track area | | | | | | | | |
|---|---|---|---|---|---|---|---|---|
| Line 1 | 1 | 1 | 1 | 1 | 1 | 1 | 1 | 0.542 |
| Line 2 | 1 | 0.985 | 0.959 | 0.941 | 0.882 | 0.831 | 0.781 | 0.371 |
| Line 3 | 0.979 | 0.906 | 0.628 | 0.624 | 0.402 | 0.545 | 0.310 | 0.249 |
| **– – continued** | | | | | | | | |
| Line 1 | 0.337 | 0.274 | 0.198 | 0.170 | 0.167 | 0.138 | 0.119 | 0.106 |
| Line 2 | 0.218 | 0.149 | 0.117 | 0.099 | 0.101 | 0.099 | 0.081 | 0.060 |
| Line 3 | 0.214 | 0.196 | 0.179 | 0.167 | 0.155 | 0.121 | 0.107 | 0.085 |
| **– – continued** | | | | | | | | |
| Line 1 | 0.087 | 0.081 | 0.045 | 0.043 | 0.037 | 0.028 | 0.018 | 0.015 |
| Line 2 | 0.047 | 0.040 | 0.021 | 0.019 | 0.022 | 0.020 | 0.007 | 0.006 |
| Line 3 | 0.072 | 0.061 | 0.050 | 0.053 | 0.044 | 0.027 | 0.017 | 0.007 |

Table 9: **Molten liquid volume fraction in the melt track area in the melt track region, re-melt region and reheat region.** The liquid volume fraction values are calculated from the peak intensity measured from the liquid phase reflections.

| Residual stress calculation - Melt track area - $\sigma_{xx}$ (MPa) | | | | | | | | |
|---|---|---|---|---|---|---|---|---|
| Line 1 | 0 | 0 | 0 | 0 | 0 | 0 | 15.514 | -2.282 |
| Line 2 | 0 | 0 | 0 | 0 | 0 | 0 | -8.390 | 2.468 |
| Line 3 | 0 | 0 | 0 | 0 | 0 | 0 | 2.379 | 18.002 |
| **– – continued** | | | | | | | | |
| Line 1 | 8.358 | 10.937 | 11.452 | 17.813 | 4.754 | 20.450 | 8.937 | 1.171 |
| Line 2 | 0.212 | 13.788 | 9.194 | -8.294 | 6.290 | -6.790 | -3.857 | 1.325 |
| Line 3 | 3.083 | -15.395 | -10.289 | -14.710 | 10.825 | -19.130 | -18.189 | -4.706 |
| **– – continued** | | | | | | | | |
| Line 1 | -0.844 | -8.367 | 48.323 | -12.443 | -0.076 | -25.783 | -46.942 | -35.201 |
| Line 2 | 8.653 | 2.413 | 41.281 | -32.129 | -22.603 | -61.874 | -54.722 | -36.805 |
| Line 3 | -18.333 | -53.708 | -24.082 | -82.943 | -77.601 | -85.757 | -67.295 | -98.869 |
| **Residual stress calculation - Melt track area - $\sigma_{xy}$ (MPa)** | | | | | | | | |
| Line 1 | 0 | 0 | 0 | 0 | 0 | 0 | -10.508 | -2.827 |
| Continued on next page | | | | | | | | |

Table 10– – continued from previous page

| | | | | | | | | |
|---|---|---|---|---|---|---|---|---|
| Line 2 | 0 | 0 0 | 0 | 0 | 0 | -1.406 | -10.120 | |
| Line 3 | 0 | 0 0 | 0 | 0 | 0 | -10.020 | -10.433 | |
| **– – continued** | | | | | | | | |
| Line 1 | -5.704 | -9.054 | -7.583 | -11.777 | 4.951 | -16.794 | -19.368 | -22.554 |
| Line 2 | -10.589 | -11.773 | -23.029 | -13.728 | -48.616 | -8.322 | -26.200 | -39.201 |
| Line 3 | -17.468 | -23.291 | -19.294 | -25.423 | -37.424 | -31.552 | -46.770 | -49.952 |
| **– – continued** | | | | | | | | |
| Line 1 | -26.759 | -32.961 | 16.748 | -52.994 | -37.154 | -49.469 | -69.535 | -44.587 |
| Line 2 | -31.077 | -26.301 | -50.127 | -52.124 | -46.049 | -49.435 | -66.359 | -54.080 |
| Line 3 | -58.689 | -52.396 | -46.648 | -75.752 | -89.662 | -97.084 | -87.537 | -84.974 |
| **Residual stress calculation - Melt track area - $\sigma_{yy}$ (MPa)** | | | | | | | | |
| Line 1 | 0 | 0 | 0 | 0 | 0 | 0 | 16.112 | 1.332 |
| Line 2 | 0 | 0 | 0 | 0 | 0 | 0 | 12.851 | 0.524 |
| Line 3 | 0 | 0 | 0 | 0 | 0 | 0 | 12.317 | 22.757 |
| **– – continued** | | | | | | | | |
| Line 1 | 7.757 | 2.752 | 13.695 | 10.647 | -1.971 | 14.114 | 22.429 | 25.404 |
| Line 2 | 16.857 | 12.747 | 9.050 | -16.081 | 28.022 | 16.474 | -12.194 | 1.521 |
| Line 3 | 15.038 | 7.840 | 7.552 | 20.988 | 24.094 | 34.425 | 33.112 | 43.969 |
| **– – continued** | | | | | | | | |
| Line 1 | 33.354 | 52.903 | 45.425 | 39.642 | 64.403 | 61.438 | 80.461 | 105.719 |
| Line 2 | 21.128 | 64.163 | 34.547 | 38.944 | 47.126 | 81.973 | 102.004 | 118.289 |
| Line 3 | 59.197 | 30.508 | 79.033 | 59.936 | 80.935 | 110.191 | 118.874 | 127.116 |

Table 10: **Residual stress in the melt track area.** The residual stress values are calculated from the lattice spacing measured from the $\gamma$ reflections.

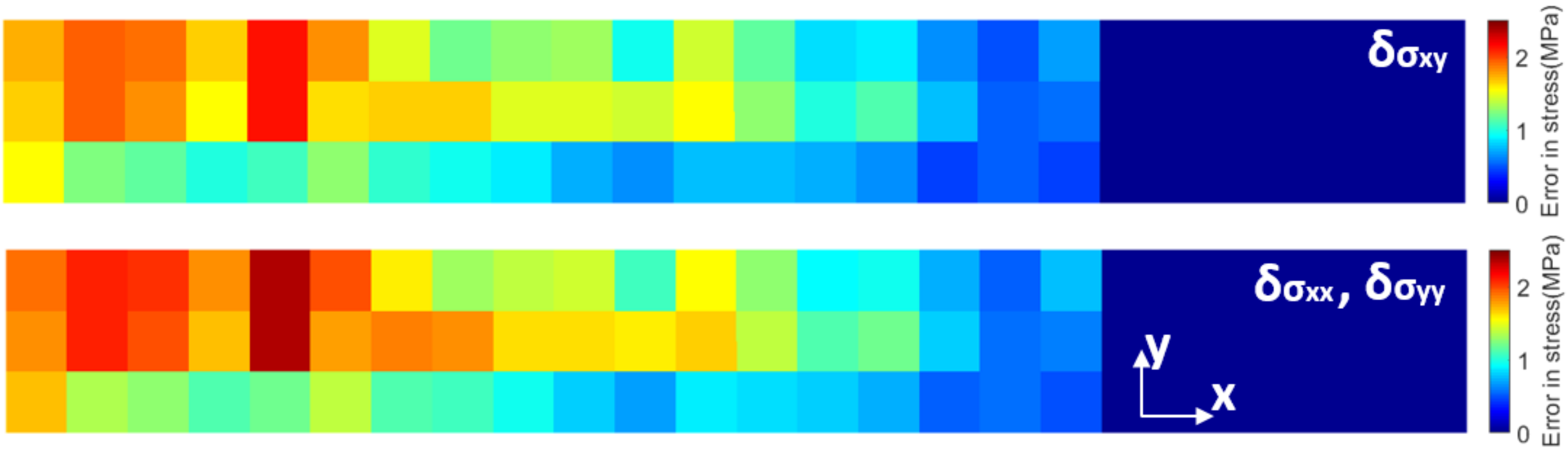

Figure 6: **Residual stress 95% confidence interval in the melt track area.** The values are calculated through error propagation from fitting error. The figure is correlated with Figure 6.

| | | | | | | | | |
|---|---|---|---|---|---|---|---|---|
| **Residual stress 95% confidence interval** | | | | | | | | |
| **$\sigma_{xx}$ 95% confidence interval** | | | | | | | | |
| Line 1 | 0 | 0 | 0 | 0 | 0 | 0 | 0.761 | 0.525 |
| Line 2 | 0 | 0 | 0 | 0 | 0 | 0 | 0.614 | 0.569 |
| Line 3 | 0 | 0 | 0 | 0 | 0 | 0 | 0.494 | 0.578 |
| **– – continued** | | | | | | | | |
| Line 1 | 0.717 | 0.947 | 0.915 | 1.267 | 1.552 | 1.065 | 1.414 | 1.376 |

Continued on next page

Table 11– – continued from previous page

| | | | | | | | | |
|---|---|---|---|---|---|---|---|---|
| Line 2 | 0.819 | 1.203 | 1.111 | 1.396 | 1.664 | 1.573 | 1.612 | 1.621 |
| Line 3 | 0.510 | 0.732 | 0.799 | 0.830 | 0.860 | 0.701 | 0.800 | 0.964 |
| **– – continued** | | | | | | | | |
| Line 1 | 1.327 | 1.587 | 1.974 | 2.344 | 1.830 | 2.068 | 2.088 | 1.886 |
| Line 2 | 1.814 | 1.843 | 1.784 | 2.355 | 1.693 | 1.971 | 2.107 | 1.814 |
| Line 3 | 1.059 | 1.125 | 1.379 | 1.197 | 1.108 | 1.251 | 1.334 | 1.703 |
| $\sigma_{xy}$ **95% confidence interval** | | | | | | | | |
| Line 1 | 0 | 0 | 0 | 0 | 0 | 0 | 0.681 | 0.477 |
| Line 2 | 0 | 0 | 0 | 0 | 0 | 0 | 0.548 | 0.510 |
| Line 3 | 0 | 0 | 0 | 0 | 0 | 0 | 0.449 | 0.524 |
| **– – continued** | | | | | | | | |
| Line 1 | 0.654 | 0.869 | 0.834 | 1.162 | 1.430 | 0.968 | 1.292 | 1.256 |
| Line 2 | 0.744 | 1.104 | 1.014 | 1.273 | 1.531 | 1.440 | 1.467 | 1.472 |
| Line 3 | 0.454 | 0.663 | 0.731 | 0.756 | 0.781 | 0.628 | 0.727 | 0.873 |
| **– – continued** | | | | | | | | |
| Line 1 | 1.207 | 1.450 | 1.811 | 2.144 | 1.679 | 1.899 | 1.917 | 1.728 |
| Line 2 | 1.652 | 1.679 | 1.621 | 2.146 | 1.545 | 1.802 | 1.939 | 1.653 |
| Line 3 | 0.961 | 1.025 | 1.254 | 1.093 | 1.007 | 1.142 | 1.221 | 1.559 |
| $\sigma_{yy}$ **95% confidence interval** | | | | | | | | |
| Line 1 | 0 | 0 | 0 | 0 | 0 | 0 | 0.761 | 0.525 |
| Line 2 | 0 | 0 | 0 | 0 | 0 | 0 | 0.614 | 0.569 |
| Line 3 | 0 | 0 | 0 | 0 | 0 | 0 | 0.494 | 0.578 |
| **– – continued** | | | | | | | | |
| Line 1 | 0.717 | 0.947 | 0.915 | 1.268 | 1.552 | 1.065 | 1.414 | 1.376 |
| Line 2 | 0.818 | 1.203 | 1.111 | 1.396 | 1.664 | 1.573 | 1.612 | 1.621 |
| Line 3 | 0.511 | 0.732 | 0.800 | 0.830 | 0.861 | 0.701 | 0.800 | 0.964 |
| **– – continued** | | | | | | | | |
| Line 1 | 1.327 | 1.587 | 1.974 | 2.344 | 1.830 | 2.068 | 2.087 | 1.885 |
| Line 2 | 1.814 | 1.843 | 1.785 | 2.355 | 1.693 | 1.971 | 2.107 | 1.814 |
| Line 3 | 1.058 | 1.125 | 1.379 | 1.197 | 1.108 | 1.250 | 1.334 | 1.702 |

Table 11: **Residual stress 95% confidence interval in the melt track area.** The values are calculated through error propagation from fitting error.

| | | | | | | | | |
|---|---|---|---|---|---|---|---|---|
| **Liquid volume fraction calculation - Melt track area** | | | | | | | | |
| Line 1 | 0.24856 | 0.30635 | 0.47351 | 0.32747 | 0.99235 | 0.18865 | 0.52987 | 0.7714 |
| Line 2 | 0.16924 | 0.21132 | 0.24706 | 0.32351 | 0.10309 | 0.43045 | 0.45941 | 0.50086 |
| Line 3 | 0.21106 | 0.26452 | 0.21753 | 0.29008 | 0.44508 | 0.36702 | 0.53193 | 0.58773 |
| **– – continued** | | | | | | | | |
| Line 1 | 0.65218 | 0.73821 | 1.1406 | 1.0421 | 0.94111 | 1.04879 | 1.40973 | 1.2897 |
| Line 2 | 0.59931 | 0.76216 | 0.79954 | 1.09194 | 0.91152 | 1.05422 | 1.45627 | 1.10896 |
| Line 3 | 0.68846 | 0.60646 | 0.59174 | 0.87706 | 1.02805 | 1.12625 | 1.03867 | 1.01138 |

Table 12: **Von Mises 95% confidnece interval in the melt track area** The values are calculated through error propagation from the fitting error. The table is correlated with Figure 7a.

## 4 Supplementary Video

**Supplementary Video 1** The Blown Powder Additive Manufacturing Process Replicator (BAMPR) has been developed to faithfully replicate a commercial DED-AM system that can be integrated in synchrotron beamlines.

**Supplementary Video 2** Time-resolved radiographs acquired during DED-AM of a multilayer thin wall melt track using IN718. IN718 build is under $P = 1.59 \times 10^3$ W mm$^{-3}$, $v = 1\,\text{mm}\,\text{s}^{-1}$, powder feedrate $1\,\text{g}\,\text{min}^{-1}$, captured at 200 $fps$.

**Supplementary Video 3** Time-series radiographs acquired showing W tracers revealing the melt pool flow patterns in the multi-layer DED-AM build and the flow pattern observed, showing that the melt pool shape is largely determined by the flow characteristics. IN718 build is under $P = 1.59 \times 10^3$ W mm$^{-3}$, $v = 1\,\text{mm}\,\text{s}^{-1}$, powder feedrate $1\,\text{g}\,\text{min}^{-1}$., captured at 5,000 fps.

## References

[1] M. Mostafavi, D. M. Collins, M. J. Peel, C. Reinhard, S. M. Barhli, R. Mills, M. B. Marshall, R. S. Dwyer-Joyce, and T. Connolley. Dynamic contact strain measurement by time-resolved stroboscopic energy dispersive synchrotron x-ray diffraction. *Strain*, 53(2):e12221, 2017.

[2] U. Welzel, J. Ligot, P. Lamparter, A. C. Vermeulen, and E. J. Mittemeijer. Stress analysis of polycrystalline thin films and surface regions by X-ray diffraction. *Journal of Applied Crystallography*, 38(1):1–29, Feb 2005.

[3] P.E. Aba-Perea, T. Pirling, P.J. Withers, J. Kelleher, S. Kabra, and M. Preuss. Determination of the high temperature elastic properties and diffraction elastic constants of ni-base superalloys. *Materials & Design*, 89:856 – 863, 2016.

[4] D.M. Collins, D.J. Crudden, E. Alabort, T. Connolley, and R.C. Reed. Time-resolved synchrotron diffractometry of phase transformations in high strength nickel-based superalloys. *Acta Materialia*, 94:244 – 256, 2015.

[5] R. Dinnebier and S. Billinge, editors. *Powder Diffraction Theoryand Practice*. The Royal Society of Chemistry, 2008.

[6] E. Prince, editor. *International Tables for Crystallography,Mathematical, Physical and Chemical Tables, vol. C*. Springer, 2006.

[7] C. T. Chantler. Theoretical form factor, attenuation, and scattering tabulation for z=1–92 from e=1–10 ev to e=0.4–1.0 mev. *Journal of Physical and Chemical Reference Data*, 24(1):71–643, 1995.

[8] J-O Andersson, T. Helander, L. Hglund, P. Shi, and B. Sundman. Thermo-calc & dictra, computational tools for materials science. *Calphad*, 26(2):273 – 312, 2002.

[9] A.D. Patel and Y.V. Murty. Effect of cooling rate on microstructural development in alloy 718. In E.A. Loria, editor, *Superalloys 718, 625, 706 and Various Derivatives*, pages 123–132, 2001.

[10] W. Kurz. Solidification microstructure-processing maps: Theory and application. *Advanced Engineering Materials*, 3(7):443–452, 2001.

[11] A. Dass and Moridi A. State of the art in directed energy deposition: From additive manufacturing to materials design. *Coatings*, 9(418):1–26, 2019.

[12] Q. Guo and Zhao C. In-situ full-field mapping of melt flow dynamics in laser metal additive manufacturing. *Additive Manufacturing*, 31:100939, 2020.

[13] R. N. Abdullaev, R. A. Khairulin, S. V. Stankus, and Yu. M. Kozlovskii. Density and volumetric expansion of the inconel 718 alloy in solid and liquid states. *Thermophys. Aeromech.*, 26:785–788, 2019.

[14] M. Hayashi, A. Jakobsson, T. Tanaka, and S. Seetharaman. Surface tension of the nickel-based superalloy cmsx-4. *High Temperature High Pressures*, 35/36:441–445, 2003/2004.

[15] G. Pottlacher, H. Hosaeus, E. Kaschnitz, and A. Seifter. Thermophysical properties of solid and liquidinconel 718 alloy. *Scandinavian Journal of Metallurgy*, 31(3):161–168, 2002.